\documentclass[10pt]{article}
\usepackage[cp1251]{inputenc}
\newcommand{\sss}{\scriptstyle}
\usepackage[dvips]{epsfig}   \usepackage[dvips]{changebar}
\usepackage[dvips]{graphicx}

\def\lsim{\
  \lower-1.2pt\vbox{\hbox{\rlap{$<$}\lower5pt\vbox{\hbox{$\sim$}}}}\ }
\def\gsim{\
  \lower-1.2pt\vbox{\hbox{\rlap{$>$}\lower5pt\vbox{\hbox{$\sim$}}}}\ }
   
  \usepackage{amsmath,enumerate, amsfonts, amssymb,amsthm}

\begin{document}
\title{Can a crystal be the ground state of a Bose system?}
 \author{Maksim D. Tomchenko
\bigskip \\ {\small Bogolyubov Institute for Theoretical Physics} \\
 {\small 14b, Metrolohichna Str., Kyiv 03143, Ukraine} }
 \date{\empty}
 \maketitle
 \large
 \sloppy
 \textit{It is usually assumed that the Bose crystal at $T=0$
corresponds to the genuine ground state of a Bose system, i.e., this
state is non-degenerate and is described by the wave function
without nodes. By means of symmetry analysis we show that the ground
state of a Bose system of any density should correspond to a liquid
or gas, but not to a crystal. The main point is that any anisotropic
state of a system of spinless bosons  is degenerate. We prove this
for an infinite three-dimensional (3D) system and a finite
ball-shaped 3D system. One can expect that it is true also for a
finite system of any form.  Therefore, the anisotropic state cannot
be the genuine ground state. Hence, a zero-temperature natural 3D
crystal should correspond to an excited state of a Bose system. The
wave function $\Psi^{c}_{0}$ of a zero-temperature 3D Bose crystal
is proposed for zero boundary conditions. Apparently, such
$\Psi^{c}_{0}$ corresponds to a local minimum of energy (absolute
minimum corresponds to a liquid). Those properties yield the
possibility of existence  of superfluid liquid H$_{2}$, Ne, Ar, and
other inert elements. We propose several possible experimental ways
of obtaining them.}


 \section{Introduction}
In the Nature, liquids usually crystallize at the cooling. This
leads to the natural commonly accepted assumption that the lowest
state of a dense three-dimensional (3D) Bose system corresponds to a
crystal. However, we will see in what follows that this is
apparently not the case. The question about the structure of the
ground state (GS) is of primary importance. In a strange way, it has
been little investigated in the literature. Below, we will try to
clarify this question mathematically (Sect. 2) and consider the
possible experimental consequences (Sect. 3). In this regard, we
mention the book by K. Mendelssohn \cite{mendelssohn1}, that
provides an excellent  review of the history of the development of
low-temperature physics till 1965.

 \section{Mathematical substantiation}
\subsection{Ans\"{a}tze for the wave function of the ground state of a Bose crystal}
Consider $N$ spinless interacting bosons without any external field.
The Hamiltonian of such a system reads
\begin{equation}
 \hat{H} = -\frac{\hbar^{2}}{2m}\sum\limits_{j=1}^{N}\triangle_{\textbf{r}_{j}} + \frac{1}{2}\sum\limits_{jl}^{l \not= j}
 U(|\textbf{r}_{l}-\textbf{r}_{j}|).
     \label{1-ham1} \end{equation}
In the literature, three solutions were proposed for the GS wave
function (WF) $\Psi^{c}_{0}$  of a Bose crystal. All of them
correspond to WF without nodes. Thus, it was assumed that the
crystal at $T=0$ corresponds to the genuine GS of the system. At
first the following localized ansatz was considered (see works
\cite{bernardes1960,saunders1962,nosanow1962,brueckner1965,nosanow1966,levesque1968,cep1993}
and reviews \cite{guyer,leggett2006,cazorla2017}):
 \begin{equation}
   \Psi_{0}^{c} \approx  e^{S_{0}}\sum\limits_{P_{c}}\prod\limits_{j=1}^{N} \varphi(\textbf{r}_{j}-\textbf{R}_{j}),
 \label{1-1}    \end{equation}
where $\textbf{r}_j$ and $\textbf{R}_j$ are the coordinates of atoms
and lattice sites, respectively, $P_{c}$ means all possible
permutations of coordinates $\textbf{r}_{j}$. In all formulae for
$\Psi(\textbf{r}_{1},\ldots,\textbf{r}_{N})$ we omit the
normalization constants.  The function $S_{0}$ is usually written in
the Bijl--Jastrow approximation \cite{bijl,mott,dingle,jastrow1955}:
 \begin{equation}
   S_{0} =  \frac{1}{2}\sum\limits_{i \neq
   j}S_{2}(\textbf{r}_{i}-\textbf{r}_{j}).
 \label{1-2}    \end{equation}
The exact formula for $S_{0}$ is as follows \cite{woo1972,feenberg1974}:
\begin{eqnarray}
S_{0}(\textbf{r}_{1},\ldots,\textbf{r}_{N}) =
\frac{1}{2!}\sum\limits^{j_{1}\neq
j_{2}}_{j_{1}j_{2}}S_{2}(\textbf{r}_{j_{1}}-\textbf{r}_{j_{2}})+
\frac{1}{3!}\sum\limits^{j_{1}\neq j_{2},j_{3}; j_{2}\neq
j_{3}}_{j_{1}j_{2}j_{3}}S_{3}(\textbf{r}_{j_{1}}-\textbf{r}_{j_{2}},\textbf{r}_{j_{2}}-\textbf{r}_{j_{3}})+\ldots
\nonumber \\ +\frac{1}{N!}\sum\limits^{j_{1}\neq
j_{2},\ldots,j_{N};\ldots; j_{N}\neq
j_{1},\ldots,j_{N-1}}_{j_{1}j_{2}\ldots
j_{N}}S_{N}(\textbf{r}_{j_{1}}-\textbf{r}_{j_{2}},\textbf{r}_{j_{2}}-\textbf{r}_{j_{3}},\ldots
, \textbf{r}_{j_{N-1}}-\textbf{r}_{j_{N}}).
 \label{1-2full}    \end{eqnarray}
Here, the sum including $S_{j}$ describes the $j$-particle
correlations. In ansatz (\ref{1-1}), the crystal lattice is
postulated, and it is assumed that the atoms execute small
oscillations near the sites. The function $\varphi(\textbf{r})$ from
(\ref{1-1}) in the approximation of small oscillations is
$\varphi(\textbf{r})=e^{-\alpha^{2} r^{2}/2}$
\cite{bernardes1960,saunders1962,nosanow1962,brueckner1965,nosanow1966,levesque1968,cep1993,guyer}.
The simple analysis shows that, for such solution, no condensate of
atoms is present \cite{leggett2006,penronz1956,prokofev2007}.

Later on, a wave ansatz was proposed \cite{woo1976,cep1976,cep1978}:
 \begin{equation}
   \Psi_{0}^{c} \approx  e^{S_{0}}e^{-\sum\limits_{j=1}^{N} \theta(\textbf{r}_{j})},
 \label{1-4}    \end{equation}
where function $\theta(\textbf{r})$ is periodic with periods of the
crystal. This solution is of the wave type and is characterized by a
condensate with WF $\Psi_{c}(\textbf{r})\simeq
e^{-\theta(\textbf{r})}$. The crystal-like solutions with a
condensate were considered in other approaches as well
\cite{gross1958,gross1960,luca,coniglio1969,kirz,nep,shlyapa2015,andreev2017,mt2020,fil2020}.

The third possible ansatz for GS of a crystal is as follows
\cite{mcmillan1965,chester1970,reatto1995,whitlock2006}:
\begin{equation}
\Psi_{0}^{c} =  e^{S_{0}}, \label{1-5}    \end{equation}
\begin{eqnarray}
S_{0}(\textbf{r}_{1},\ldots,\textbf{r}_{N}|\textbf{k}_{r}) =
\frac{1}{2!}\sum\limits^{j_{1}\neq
j_{2}}_{j_{1}j_{2}}S_{2}(\textbf{r}_{j_{1}}-\textbf{r}_{j_{2}}|\textbf{k}_{r})+
\frac{1}{3!}\sum\limits^{j_{1}\neq j_{2},j_{3}; j_{2}\neq
j_{3}}_{j_{1}j_{2}j_{3}}S_{3}(\textbf{r}_{j_{1}}-\textbf{r}_{j_{2}},\textbf{r}_{j_{2}}-\textbf{r}_{j_{3}}|\textbf{k}_{r})+\ldots
\nonumber \\ +\frac{1}{N!}\sum\limits^{j_{1}\neq j_{2},\ldots,j_{N};
\ldots ;j_{N}\neq j_{1},\ldots,j_{N-1}}_{j_{1}j_{2}\ldots
j_{N}}S_{N}(\textbf{r}_{j_{1}}-\textbf{r}_{j_{2}},\textbf{r}_{j_{2}}-\textbf{r}_{j_{3}},\ldots
, \textbf{r}_{j_{N-1}}-\textbf{r}_{j_{N}}|\textbf{k}_{r}).
 \label{1-5b}    \end{eqnarray}
It is a translationally invariant anisotropic solution. We denote
the anisotropy of function (\ref{1-5b}) by a vector $\textbf{k}_{r}$
(this is the reciprocal lattice vector with the nonzero smallest
components). It is known that GS of a liquid or a gas is described
by an isotropic WF (\ref{1-5}), (\ref{1-2full})
\cite{jastrow1955,woo1972,feenberg1974,bz1956,fey1972,yuv1} (we
consider the function (\ref{1-2full}) to be isotropic). It was
assumed in a number of works that, at some critical density
$\rho=\rho_{c},$ the liquid solution (\ref{1-5}), (\ref{1-2full})
spontaneously transforms into a crystalline solution  (\ref{1-5}),
(\ref{1-5b}) \cite{mcmillan1965,chester1970,reatto1995,reatto2009}.

Let us clarify which of functions (\ref{1-1}), (\ref{1-4}) and
(\ref{1-5}), (\ref{1-5b}) can be the solution for a crystal. In
order to verify the bulk structure of solutions, we can use any
boundary conditions (BCs). Let us test the crystal solutions
(\ref{1-1}), (\ref{1-4}), and (\ref{1-5}), (\ref{1-5b}) for periodic
BCs. The periodic system is translationally invariant, which yields
two consequences. (i) The properties of a system on a ring must not
change at a rotation of the ring. This holds provided that, at a
displacement of the system as a whole by the radius-vector $\delta
\textbf{r}\rightarrow \textbf{0}$,  WF of the system is multiplied
by a constant:
\begin{equation}
\Psi(\textbf{r}_{1}+\delta \textbf{r},\ldots,\textbf{r}_{N}+\delta
\textbf{r}) = (1+i \textbf{p}\delta
\textbf{r})\Psi(\textbf{r}_{1},\ldots,\textbf{r}_{N})= e^{i
\textbf{p}\delta
\textbf{r}}\Psi(\textbf{r}_{1},\ldots,\textbf{r}_{N}).
 \label{1-6} \end{equation}
(ii) Since
\begin{equation}
\Psi(\textbf{r}_{1}+\delta \textbf{r},\ldots,\textbf{r}_{N}+\delta
\textbf{r}) = \left (1+\delta
\textbf{r}\sum\limits_{j}\frac{\partial}{\partial
\textbf{r}_{j}}\right)\Psi(\textbf{r}_{1},\ldots,\textbf{r}_{N} ),
 \label{1-6b} \end{equation}
relation (\ref{1-6}) yields
\begin{equation}
\hat{\textbf{P}}\Psi\equiv -i\hbar
\sum\limits_{j}\frac{\partial}{\partial \textbf{r}_{j}}\Psi= \hbar
\textbf{p}\Psi.
 \label{1-77}    \end{equation}
Therefore, the full collection of WFs
$\Psi_{j}(\textbf{r}_{1},\ldots,\textbf{r}_{N})$ of such a
boundary-value problem can be constructed so that each WF is an
eigenfunction of the momentum operator  $\hat{\textbf{P}}$, i.e., it
satisfies conditions (\ref{1-6}) and (\ref{1-77}).
This is well known from quantum mechanics.

The most widely used ansatz is WF (\ref{1-1}), where the coordinates
of sites $\textbf{R}_{j}$ are \textit{fixed} and the same ones at
any possible values of the atomic coordinates $\{\textbf{r}_{j}\}$
(including the sets $\{\textbf{r}_{j}\}$  and
$\{\textbf{r}_{j}+\textbf{a}\}$)
\cite{saunders1962,brueckner1965,nosanow1966,guyer,leggett2006,cazorla2017}.
Such ansatz does not satisfy conditions (\ref{1-6}) and
(\ref{1-77}). Indeed, for $\varphi(\textbf{r})=e^{-\alpha^{2}
r^{2}/2}$ we have
\begin{equation}
\hat{\textbf{P}}\Psi_{0}^{c}=  i\hbar \alpha^{2} \Psi_{0}^{c}
\sum\limits_{j}(\textbf{r}_{j}-\textbf{R}_{j}) \neq \hbar
\textbf{p}\Psi_{0}^{c}.
 \label{1-8}    \end{equation}
With regard for the anharmonic corrections to $\varphi(\textbf{r}),$
the formula for $\hat{\textbf{P}}\Psi_{0}^{c}$ is complicated, but
the conclusion does not change. More complicated modification of WF
(\ref{1-1}) was proposed  in \cite{koehler1966}. For it, relation
(\ref{1-77}) does not hold as well.

Solution (\ref{1-1}) is impossible for periodic BCs also because the
concentration of a periodic Bose system is an exact constant:
$n(\textbf{r})=const$ \cite{mt2020,vak1989,vak1990,sacha2018}. This
surprising property is related to the translation invariance and can
be easily proved (for any pure state with a definite momentum,
including the lowest state ($T=0$), see the calculation of the
density matrix in the coordinate representation in \cite{sacha2018}
and in the operator approach in \cite{mt2020}; for $T>0,$ this can
be proved analogously to the analysis in \cite{sacha2018}, using the
formula $n(\textbf{r})=const\cdot\int d\textbf{r}_{2}\ldots
d\textbf{r}_{N}\sum_{j}e^{-E_{j}/k_{B}T}
|\Psi_{j}(\textbf{r},\textbf{r}_{2},\ldots,\textbf{r}_{N})|^{2}$ and
property (\ref{1-6})).  The constancy of the density means that, in
a periodic system, the crystalline ordering is hidden. It must
manifest itself in oscillations (with the period  of a crystal) of
the two-particle density matrix
$g_{2}(\textbf{r}_{1},\textbf{r}_{2})$, rather than in the density.
But solution (\ref{1-1}) corresponds exactly to the oscillating
particle \textit{density}: $n(\textbf{r}+\textbf{b})=n(\textbf{r})$,
where $b_{x},b_{y},b_{z}$ are the sizes of crystal cell. [Let us
show it. Since $S_{0}$ in (\ref{1-2}) and (\ref{1-2full}) correspond
to a constant density, we set $S_{0}=0$ in (\ref{1-1}). Then we get
$n(\textbf{r})= \tilde{C} \cdot \sum_{j}e^{-\alpha^{2}
(\textbf{r}-\textbf{R}_{j})^{2}} \neq const$. On the other hand,
$n(\textbf{r}+\textbf{b})=n(\textbf{r})$, since the translation of
the crystal by one step is equivalent to the renumbering of sites,
which does not change the sum.]

For the wave ansatz (\ref{1-4}), we obtain
\begin{equation}
\hat{\textbf{P}}\Psi_{0}^{c}=  \Psi_{0}^{c} i\hbar
\sum\limits_{j}\frac{\partial \theta(\textbf{r}_{j})}{\partial
\textbf{r}_{j}}.
 \label{1-8c}  \end{equation}
This equals $\hbar \textbf{p}\Psi_{0}^{c}$, if
$\theta(\textbf{r}_{j})=-i\textbf{p}\textbf{r}_{j}/N+const$. The
ground state must correspond to $\hbar\textbf{p}=0$ (see below). We
get $\textbf{p}=0$ at $\theta(\textbf{r}_{j})=const$. Then ansatz
(\ref{1-4}) is reduced to the solution (\ref{1-5}) with $S_{0}$
(\ref{1-2full}). However, it is a solution for WF of the ground
state of a \textit{uniform and isotropic system} (liquid or gas)
\cite{jastrow1955,woo1972,feenberg1974,bz1956,fey1972,yuv1}.

It is important that WF of the ground state of a crystal (or liquid)
with periodic BCs
must correspond to \textit{zero momentum}: $\hbar\textbf{p}=0$.
The case $\textbf{p}\neq 0$ is unphysical. Indeed, if the crystal
would contain a  quasiparticle, then the momentum
$\hbar\textbf{p}\neq 0$ would be associated with this quasiparticle.
But the ground state contains no quasiparticles, and the crystal as
a whole does not move. Therefore, the separated direction that is
set by nonzero momentum cannot be associated with a physical
property. In view of this, we have $\textbf{p}=0$.

This can be shown mathematically. Hamiltonian (\ref{1-ham1})
commutes with the operator of inversion $\hat{I}$ and the operator
of total momentum $\hat{\textbf{P}}$, but two last operators do not
commute with each other. According to the well-known theorem of
quantum mechanics \cite{land3,vak}, each energy level of such a
system should be degenerate. The exception is the energy level
corresponding to the zero momentum. Indeed, let the state $\Psi_{1}$
correspond to the momentum $\hbar\textbf{p}_{1}$ and the energy
$E_{1}$: $\hat{\textbf{P}}\Psi_{1}=\hbar\textbf{p}_{1}\Psi_{1}$ and
$\hat{H}\Psi_{1}=E_{1}\Psi_{1}$. Let us act by the inversion
operator on the equation $\hat{H}\Psi_{1}=E_{1}\Psi_{1}$. Since
$\hat{I}\hat{H}- \hat{H}\hat{I}=0$, we get
$\hat{H}\hat{I}\Psi_{1}=E_{1}\hat{I}\Psi_{1}$. That is, the state
$\hat{I}\Psi_{1}(\textbf{r}_{1},\ldots,\textbf{r}_{N})\equiv\Psi_{1}(-\textbf{r}_{1},\ldots,-\textbf{r}_{N})=\Psi_{2}(\textbf{r}_{1},\ldots,\textbf{r}_{N})$
corresponds to the same energy as that of the state
$\Psi_{1}(\textbf{r}_{1},\ldots,\textbf{r}_{N})$. If $\Psi_{1}$
satisfies periodic BCs, then $\Psi_{2}$ satisfies the same ones. On
the other hand, if
$\hat{\textbf{P}}\Psi_{1}(\textbf{r}_{1},\ldots,\textbf{r}_{N})=\hbar\textbf{p}_{1}\Psi_{1}(\textbf{r}_{1},\ldots,\textbf{r}_{N})$,
then
$\hat{\textbf{P}}\Psi_{2}(\textbf{r}_{1},\ldots,\textbf{r}_{N})\equiv\hat{\textbf{P}}\Psi_{1}(-\textbf{r}_{1},\ldots,-\textbf{r}_{N})
=-\hbar\textbf{p}_{1}\Psi_{1}(-\textbf{r}_{1},\ldots,-\textbf{r}_{N})=-\hbar\textbf{p}_{1}\Psi_{2}(\textbf{r}_{1},\ldots,\textbf{r}_{N})$.
In other words, the state $\Psi_{2}$ is characterized by the
momentum $-\hbar\textbf{p}_{1}$. Since the states $\Psi_{1}$ and
$\Psi_{2}$ correspond to different momenta, they are physically
different states. Hence, the level $E_{1}$ is degenerate. Only at
$\textbf{p}_{1}=0$ the states $\Psi_{1}$ and $\Psi_{2}$ are
characterized by  the identical energies and momenta. Such states
are equivalent. In this case, the level $E_{1}$ is non-degenerate.
If WF $\Psi_{1}$ corresponds to the energy $E_{1}$ and does not
correspond to a definite momentum, it can be expanded in WFs of
states with a definite momentum and the energy $E_{1}$. Such
expansion must contain at least one WF with $\hbar\textbf{p}\neq 0$,
i.e., the state $\Psi_{1}$ is degenerate. However, GS must  be
non-degenerate \cite{gilbert}. Thus, the genuine GS can correspond
only to zero momentum.

The structure of WF of the ground state, $\Psi_{0}$, of a Bose
system with periodic BCs can be easily determined. The condition
$\textbf{p}=0$  and formulae (\ref{1-6}) and (\ref{1-77}) imply that
$\Psi_{0}$ should not vary at a translation. Therefore, it can
depend only on the difference of coordinates. The general form of
such nodeless function is given by formulae (\ref{1-5}),
(\ref{1-2full}) or (\ref{1-5}), (\ref{1-5b}). This can be proved
strictly (see \cite{yuv1} and Appendix 1 below).


It is interesting to note that the structure of $\Psi_{0}$
(\ref{1-5}), (\ref{1-2full}) of a quantum liquid is usually obtained
from the requirement that $\Psi_{0}$ should be invariable at a
translation [$\textbf{p}=0$ in (\ref{1-6})] \cite{woo1972,yuv1}.
However, the translation invariance of a system admits
$\textbf{p}\neq 0$ in (\ref{1-6}). Apparently, it was not shown
previously in the literature that the Bose system with
$\textbf{p}\neq 0$ does not correspond to GS. This was proved above.
Therefore, the condition $\textbf{p}=0$ is primary, and the
translation invariance of $\Psi_{0}$ is a consequence of this
requirement.

Note also that, in a series of articles
\cite{reatto1995,reatto2009,vitiello1988,reatto1998,reatto2011},
Bose liquid  and Bose crystal were described by a ``shadow'' WF
(SWF)
 \begin{equation}
\Psi(R)=\int e^{-\Xi(R,S)}dS, \quad
R\equiv\textbf{r}_{1},\ldots,\textbf{r}_{N},
 \label{1-s1}    \end{equation}
\begin{equation}
\Xi(R,S)=\sum_{j_{1}<j_{2}}u_{r}(|\textbf{r}_{j_{1}}-\textbf{r}_{j_{2}}|)+
\sum_{k}u_{sr}(|\textbf{r}_{k}-\textbf{s}_{k}|)+\sum_{j_{3}<j_{4}}u_{s}(|\textbf{s}_{j_{3}}-\textbf{s}_{j_{4}}|),
 \label{1-s2}    \end{equation}
where $S\equiv\textbf{s}_{1},\ldots,\textbf{s}_{N}$ are ``shadow''
variables. If $u_{r}, u_{sr}, u_{s}$ are represented in the form of
Fourier series, then SWF (\ref{1-s1}), (\ref{1-s2}) becomes
translationally invariant. Since any nodeless translationally
invariant WF can be written in the form of (\ref{1-5}),
(\ref{1-2full}) or (\ref{1-5}), (\ref{1-5b}), function (\ref{1-s1}),
(\ref{1-s2}) is a partial case of the general solution (\ref{1-5}),
(\ref{1-2full}) [or (\ref{1-5}), (\ref{1-5b})] written in a
different form. This was noted in reviews
\cite{reatto1995,whitlock2006}. In this case, SWF has a relatively
simple structure and, apparently, enables one to indirectly involve,
at least partially, all higher correlation corrections
\cite{reatto1995}. The authors of works
\cite{reatto1995,reatto2009,vitiello1988,reatto1998,reatto2011}
assert that SWF (\ref{1-s1}), (\ref{1-s2}) describes a Bose liquid
at some densities and a Bose crystal at other ones. This is
equivalent to the assumption that, at some $\rho=\rho_{c}$, the
liquid solution (\ref{1-5}), (\ref{1-2full})  spontaneously
transforms into a crystalline solution (\ref{1-5}), (\ref{1-5b}).

We have shown above that the ground state of a periodic system of
interacting spinless bosons corresponds to zero momentum. Among four
above-considered solutions (liquid solution and three crystalline
ones), this requirement is satisfied by the liquid solution
(\ref{1-5}), (\ref{1-2full}) and the crystalline one (\ref{1-5}),
(\ref{1-5b}). The other two crystalline solutions, (\ref{1-1}) and
(\ref{1-4}), describe the states without a definite momentum.
\textit{Hence, only WF (\ref{1-5}), (\ref{1-5b}) can be an exact
solution for GS of a crystal with periodic BCs. The key question is
as follows: Can such crystalline solution exist?}

\subsection{Degeneracy of an anisotropic state}
The available literature gives no answer to the above question, to
our knowledge. Apparently, it is impossible to find it analytically.
The numerical methods also do not give an exact answer because they
give incomplete information. However, the answer can be found by
means of the symmetry analysis considered in what follows. Two other
methods are proposed in \cite{mtcrystal2,mt1Dcrystal}.

Hamiltonian (\ref{1-ham1}) of the 3D system is invariant under the
group of  orthogonal transformations $O(3)=SO(3)\times C_{i}$
consisting of the group of rotations $SO(3)$ and the group of
inversions $C_{i}$ (it contains two elements: inversion $I$ and
$I^{2}=1$; here and below, we consider that any symmetry
transformation is applied to all coordinates
$\textbf{r}_{1},\ldots,\textbf{r}_{N}$). This invariance is
preserved if $\hat{H}$ also includes a three-particle interaction
\cite{axilrod1943,bruch1973,loubeyre1988,boronat1994}. Therefore,
$\hat{H}$ commutes with the operator of rotation
$\hat{R}=e^{i\varphi \textbf{i}_{\varphi}\hat{\textbf{L}}/\hbar}$
\cite{land3,vak,petrashen}, where $\hat{\textbf{L}}$ is the operator
of total angular momentum of the system, $\varphi$ is a rotation
angle, and the unit vector $\textbf{i}_{\varphi}$ sets the rotation
axis. Hence, $[\hat{H},\hat{\textbf{L}}]=0,$ and
$[\hat{H},\hat{\textbf{L}}^{2}]=0$. Since the operators $\hat{H}$,
$\hat{\textbf{L}}^{2}$ and $\hat{L}_{z}$ commute with each other,
the complete set of eigenfunctions can be constructed so that those
functions be the eigenfunctions of these three operators
\cite{land3,vak}. It is important that BCs must admit this (as a
rule, this point is not mentioned in textbooks). Therefore, we
consider the system to be infinite (closed or not) or finite
ball-shaped. In the first case, BCs are invariant under the
translations and rotations, which corresponds to the uniformity and
isotropy of the space. This gives the laws of conservation of the
momentum and angular momentum \cite{land3,vak}. If BCs would not be
uniform and isotropic at infinity, then the laws of conservation of
the momentum and angular momentum would not hold in our world. For a
finite ball, BCs are invariant only under the rotations. In both
cases, since the Hamiltonian and BCs are invariant with respect to
the rotations, WFs can be set so that they are transformed by the
irreducible representations of the rotation group $SO(3)$
\cite{petrashen,elliott}. This group is characterized by the
complete collection of irreducible representations $g\rightarrow
\hat{T}_{l}(g)$ with $l=0, 1/2, 1, 3/2, 2, \ldots, \infty $ and  the
dimension $2l+1$ (here, $g$ is an element of the group). The
representations with integer and half-integer $l$ are, respectively,
one- and two-valued \cite{petrashen,elliott,gelfand,golod}. The
scalar WFs can be transformed only by the representations with
integer $l$. In this case, $\hat{\textbf{L}}^{2}\Psi^{(l)}=\hbar^{2}
l(l+1)\Psi^{(l)}$ \cite{petrashen,elliott}. The operator of rotation
is defined by the formula \cite{land3,vak}
\begin{equation}
 \hat{R}\Psi(\textbf{r}_{1},\ldots,\textbf{r}_{N}|\textbf{k}) = \Psi(\acute{\textbf{r}}_{1},\ldots,\acute{\textbf{r}}_{N}|\textbf{k}),
     \label{1-new0} \end{equation}
where $\textbf{r}_{j}$ and $\acute{\textbf{r}}_{j}=A\textbf{r}_{j}$
are the coordinates of a vector before and after a rotation, $A$ is
the rotation matrix (the vectors $\textbf{r}_{j}$ and
$\acute{\textbf{r}}_{j}$ are set in the same basis). The operator
$\hat{R}$ transforms the coordinates and does not affect constants
like $\textbf{k}$ that characterize a possible anisotropy of the
system.  Eq. (\ref{1-new0}) leads  to the formula
$\hat{R}=e^{i\varphi \textbf{i}_{\varphi}\hat{\textbf{L}}/\hbar}$
\cite{land3,vak}. The wave functions invariable relative to any
rotation ($\hat{R}\Psi \equiv e^{i\varphi
\textbf{i}_{\varphi}\hat{\textbf{L}}/\hbar}\Psi =\Psi$, i.e.,
$\hat{\textbf{L}}\Psi=0$) are transformed by the identical (unit)
representation $g\rightarrow\hat{T}_{0}(g)$:
$\hat{T}_{0}(g)\Psi=\Psi$ for any element  $g$ of the group $SO(3)$.
The crystalline GS is anisotropic. Therefore, $\hat{R}\Psi \neq
\Psi$, $\hat{\textbf{L}}\Psi \neq 0$. Such state is transformed by
one or several representations $g\rightarrow\hat{T}_{l}(g)$ with
$l\neq 0$. The irreducible representation
$g\rightarrow\hat{T}_{l}(g)$ is characterized by the orthonormalized
basis $\Psi^{(l)}_{1},\Psi^{(l)}_{2},\ldots,\Psi^{(l)}_{2l+1}$. In
this case,
$\hat{T}_{l}(g)\Psi^{(l)}_{j}=\sum_{p=1}^{2l+1}T^{(l)}_{pj}(g)\Psi^{(l)}_{p}$
for any element  $g$ of the group $SO(3)$, where $T^{(l)}(g)$ are
the matrices of constants and realize the representation
$g\rightarrow\hat{T}_{l}(g)$ \cite{petrashen,elliott}. For each
representation $g\rightarrow\hat{T}_{l}(g),$ all functions
correspond to the same energy. Indeed, let WF $\Psi^{(l)}_{j}$ be an
eigenfunction of the Schr\"{o}dinger equation with energy $E$:
\begin{equation}
 \hat{H}\Psi^{(l)}_{j} = E\Psi^{(l)}_{j}.
     \label{1-new00} \end{equation}
Let us act by the operator
$\hat{T}(g)=\hat{R}^{-1}(g)=\hat{R}(g^{-1})$ \cite{elliott,gelfand}
on this equation. Since $\hat{R}(g)\hat{H}-\hat{H}\hat{R}(g)=0$ for
any rotation $g$, the last equality holds also for the rotation
$g^{-1}$. From whence, we get
$\hat{T}(g)\hat{H}-\hat{H}\hat{T}(g)=0$. Therefore,
\begin{equation}
 E\hat{T}(g)\Psi^{(l)}_{j}=\hat{T}(g)E\Psi^{(l)}_{j}=\hat{T}(g)\hat{H}\Psi^{(l)}_{j}
=\hat{H}\hat{T}(g)\Psi^{(l)}_{j}.
     \label{1-new000} \end{equation}
That is, the function $\hat{T}(g)\Psi^{(l)}_{j}$ is also an
eigenfunction of the Schr\"{o}dinger equation with energy $E$. We
now substitute the expansion
$\hat{T}(g)\Psi^{(l)}_{j}\equiv\sum_{p}\hat{T}_{p}(g)\Psi^{(l)}_{j}=\hat{T}_{l}(g)\Psi^{(l)}_{j}=\sum_{p=1}^{2l+1}T^{(l)}_{pj}(g)\Psi^{(l)}_{p}$
in formula (\ref{1-new000}). Since the basis functions
$\Psi^{(l)}_{p}$ are independent of one another, we get that all
functions $\Psi^{(l)}_{p=1,\ldots,2l+1}$ are eigenfunctions of the
Schr\"{o}dinger equation with energy $E$.  Therefore, such state is
$(2l+1)$-fold degenerate. WF of a crystal
$\Psi(\textbf{r}_{1},\ldots,\textbf{r}_{N}|\textbf{k}_{r})$ may not
coincide with the function $\Psi^{(l)}_{p}$. Then it is necessary to
expand $\Psi(\textbf{r}_{1},\ldots,\textbf{r}_{N}|\textbf{k}_{r})$
in the basis functions $\Psi^{(l)}_{p}$ of all irreducible
representations corresponding to the energy of the crystal. In this
case, the degeneracy multiplicity is equal to the sum of the
dimensions $2l+1$ of all these representations. Thus, \textit{only
the isotropic state is not degenerate}. It is the state that
transits into itself at any rotation and is transformed by the unit
representation of the group $SO(3)$. Since the genuine GS of a Bose
system is non-degenerate \cite{gilbert} (see also Appendix 2 below),
it should correspond to an isotropic state. In Appendix 1 it is
shown that $\Psi_{0}$ (\ref{1-5}), (\ref{1-2full}) corresponds to
$\hat{\textbf{L}}\Psi_{0}=0$ and, therefore, is isotropic. Thus, we
have proved that any anisotropic state of an infinite (or finite
ball-shaped) 3D system of spinless bosons is degenerate and,
therefore, does not correspond to the genuine GS of the system.

WFs of a many-boson system are usually  constructed as
eigenfunctions of the momentum operator
\cite{fey1972,yuv1,yuv2,feenberg,woo1970,intro2002}. In this case,
WF of any excited  state of a \textit{periodic} system of $N$ bosons
can be written in the form \cite{yuv2,holes2020} (see also Appendix
1)
 \begin{equation}
\Psi_{\textbf{p}}(\textbf{r}_1,\ldots ,\textbf{r}_N) =
  \psi_{\textbf{p}}\Psi_0,
  \label{1-new0000}     \end{equation}
       \begin{eqnarray}
 \psi_{\textbf{p}} & =&
  b_{1}(\textbf{p})\rho_{-\textbf{p}} +
 \sum\limits_{\textbf{q}_{1}\neq 0}^{\textbf{q}_{1}+\textbf{p}\neq 0}
  \frac{b_{2}(\textbf{q}_{1};\textbf{p})}{2!N^{1/2}}
 \rho_{\textbf{q}_{1}}\rho_{-\textbf{q}_{1}-\textbf{p}}
 + \ldots + \nonumber \\ &+&  \sum\limits_{\textbf{q}_{1},\ldots,\textbf{q}_{N-1}\neq 0}^{\textbf{q}_{1}+\ldots +\textbf{q}_{N-1}+\textbf{p}\not= 0}
  \frac{b_{N}(\textbf{q}_{1},\ldots,\textbf{q}_{N-1};\textbf{p})}{N!N^{(N-1)/2}}
 \rho_{\textbf{q}_1}\ldots\rho_{\textbf{q}_{N-1}}
 \rho_{-\textbf{q}_{1} - \ldots - \textbf{q}_{N-1}-\textbf{p}}
        \label{psip-2}\end{eqnarray}
with $\Psi_0$ (\ref{1-5}), (\ref{1-2full}). Formulae
(\ref{1-new0000}) and (\ref{psip-2}) are exact. Solution
(\ref{1-new0000}), (\ref{psip-2}) describes states with one, two, or
many interacting phonons (or rotons), depending on the coefficients
$b_{j}$ \cite{holes2020}. WF (\ref{1-new0000}), (\ref{psip-2})
corresponds to the momentum $\hbar\textbf{p}$.  The translation
$\{\textbf{r}_{j}\}\rightarrow \{\textbf{r}_{j}+\textbf{a}\}$
transfers solution (\ref{1-new0000}) into the equivalent solution:
$\Psi_{\textbf{p}}(\textbf{r}_1,\ldots ,\textbf{r}_N)\rightarrow
e^{i\textbf{p}\textbf{a}} \Psi_{\textbf{p}}(\textbf{r}_1,\ldots
,\textbf{r}_N)$.  Such approach assumes that $\hat{H}$ is invariant
under translations and, therefore, commutes with the operator of
translations $\hat{\textbf{T}}=e^{i\textbf{a}\textbf{P}/\hbar}$ and
the operator of total momentum $\hat{\textbf{P}}$. In view of this,
one can find the complete set of orthogonal functions, being the
eigenfunctions of the operators $\hat{H}$ and $\hat{\textbf{P}}$ (it
differs from the complete set for the operators $\hat{H}$,
$\hat{\textbf{L}}^{2}$, and $\hat{L}_{z}$, since $\hat{\textbf{P}}$
and $\hat{\textbf{L}}$ do not commute with each other). Functions
(\ref{1-new0000}) with all possible $\textbf{p}$ and the function
$\Psi_0$ (\ref{1-5}), (\ref{1-2full}) realize such a set
\cite{yuv1,yuv2,holes2020}. It can be called a $\textbf{P}$-set.
Earlier, the analysis of crystalline solutions was performed only
within the $\textbf{P}$-approach. Apparently, this is why the
degeneracy of the anisotropic state was not noticed.

The many-boson systems are usually described in the
$\textbf{P}$-approach, since it is simpler and more physical (as a
rule, a quasiparticle is characterized by a definite momentum,
rather than an angular momentum). The $\textbf{L}$-set of
eigenfunctions is used for the description of the electron shell of
an atom. For many-boson systems, the $\textbf{L}$-approach was used
rarely and only for the corresponding BCs \cite{cr1,cr2}. If BCs
admit both approaches, then any WF from the $\textbf{P}$-set can be
expanded in the complete set of WFs from the $\textbf{L}$-set, and
vice versa. Interestingly, $\Psi_{0}$ (\ref{1-5}), (\ref{1-2full})
belongs simultaneously to the $\textbf{P}$- and $\textbf{L}$-set of
WFs. This function is transformed by the unit representations of the
$SO(3)$ group and the group of translations $T(3)$.

Thus, for two types of systems (finite ball-like and infinite ones),
we have shown that the crystal solutions do not contain the genuine
(nodeless) GS and, therefore, do not form the complete set of
eigenfunctions of the Schr\"{o}dinger boundary-value problem. The
genuine GS correspond always to the liquid solution. Based on the
liquid GS, one can construct a set of excited liquid states. The
group $SO(3)$ contains, as subgroups, point groups corresponding to
seven types (syngonies) of crystal lattices. Therefore, it is
obvious that, for two indicated types of systems, the complete set
of WFs of the Schr\"{o}dinger boundary-value problem contains
\textit{all solutions} for each syngony and all solutions for a
liquid. The lowest state of a crystal of each type must correspond
to WF with a large number of nodes. For the infinite system such a
solution can be rotated by any angle, then we will apparently obtain
another solution of the same boundary-value problem
\cite{mtcrystal2}. That is, the complete set of WFs of a Bose system
should contain solutions for crystals of all possible types,
including the infinite number (for the infinite system) of all
admissible rotated solutions of each type. In this case, the
collection of solutions for a crystal with fixed lattice and
orientation contains the infinite number of WFs, but is an
infinitely small part of the complete collection of WFs of the
boundary-value problem. The latter can be found by means of the
construction of the general system of eigenfunctions of the
operators $\hat{H}$, $\hat{\textbf{L}}^{2}$, and $\hat{L}_{z}$ or
(for the infinite system) the operators $\hat{H}$ and
$\hat{\textbf{P}}$. It is known that each type of crystals is stable
at definite densities. In such interval of densities, the lowest
state of a crystal must correspond to a local
statistical-thermodynamic minimum of the energy in the space of
states (this is the minimum in the sense that small perturbations of
a crystal increase its energy). Otherwise, the crystal would be
unstable. In this case, the absolute minimum corresponds to a liquid
(see Fig. 1).

\begin{figure}[ht]
\centerline{\includegraphics[width=85mm]{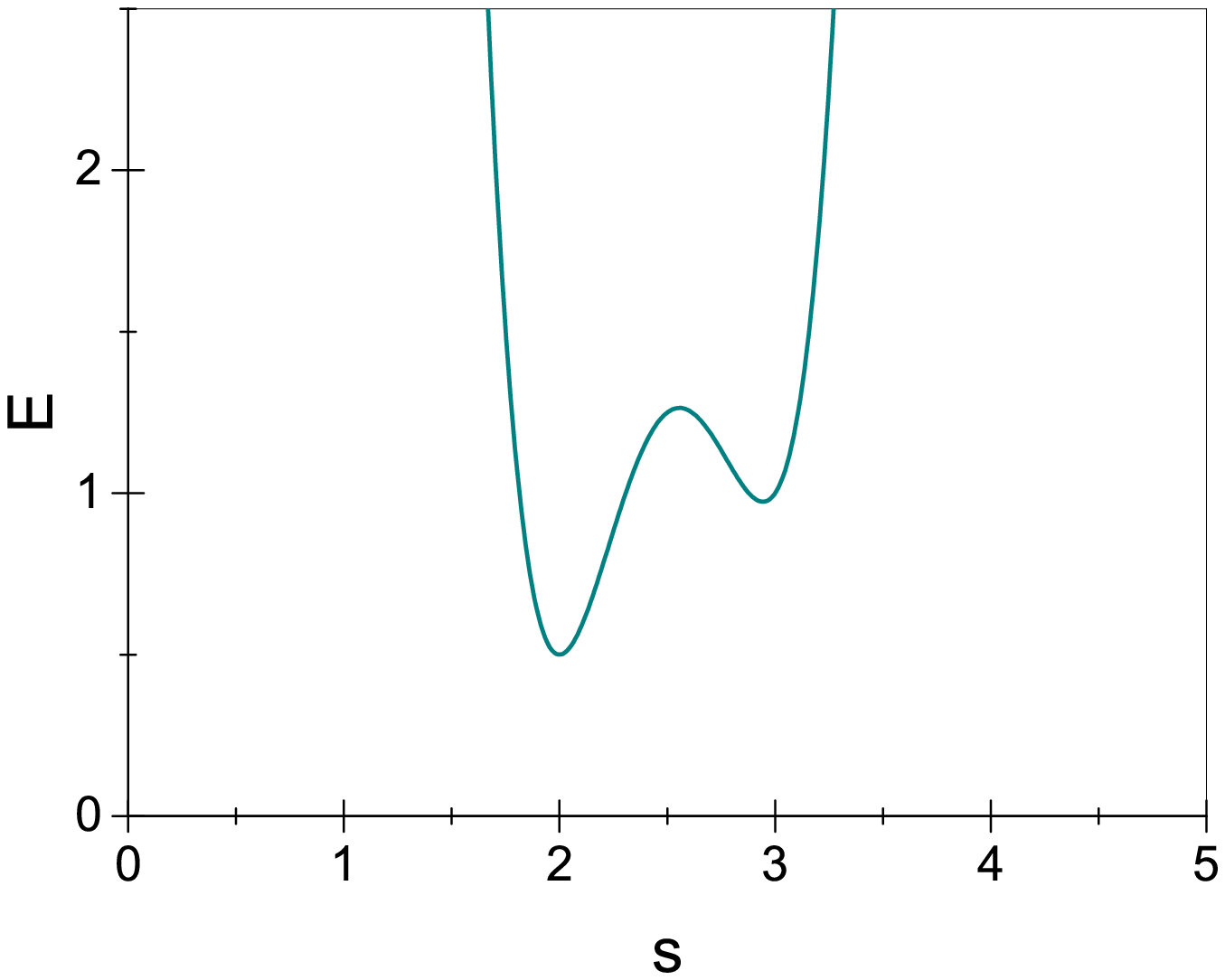} } \caption{
[Color online] The lower and upper minima correspond to a liquid and
a crystal, respectively. $E$ is the internal energy of a Bose
system, $s$ marks quantum states
$\Psi_{j}(\textbf{r}_{1},\ldots,\textbf{r}_{N})$ of the system
corresponding to the same concentration $n=N/V$. $E$ and $s$ are
given in arbitrary units.
 \label{fig1}}
\end{figure}

For clarity, consider the following example. Assume that the
Hamiltonian $\hat{H}_{c}$ of a Bose system contains an external
potential $U_{c}$ corresponding to a point symmetry group of a
crystal of some type:
\begin{equation}
 \hat{H}_{c} = -\frac{\hbar^{2}}{2m}\sum\limits_{j=1}^{N}\triangle_{\textbf{r}_{j}} + \frac{1}{2}\sum\limits_{jl}^{l \not= j}
 U(|\textbf{r}_{l}-\textbf{r}_{j}|)+ U_{c}(\textbf{r}_{1},\ldots,\textbf{r}_{N}).
     \label{1-ham2} \end{equation}
In this case, the Hamiltonian is characterized by two symmetries:
the continuous group $O(3)$ and a discrete group that is a subgroup
of the group $O(3)$ and  corresponds to the potential $U_{c}$. It is
clear that if $U_{c}$ sets a very  deep well on the place of each
lattice site, then the structure of low-lying states must be defined
by the potential $U_{c}$. That is, all WFs of low-lying states must
be transformed by irreducible representations of the discrete group
of the crystal. In this case, the crystalline solutions apparently
form the complete set of eigenfunctions of the given boundary-value
problem, and GS of the crystal should  correspond to the nodeless
non-degenerate GS of the system. As the potential $|U_{c}|$
decreases, the solutions for WFs should approach those for the
unperturbed Hamiltonian (\ref{1-ham1}). If the potential $U_{c}$ is
weak relative to the interatomic one, then the structure of all WFs
has to be determined by the group $O(3)$. Therefore, we may expect
that, at some small \textit{finite} $U_{c},$ WFs acquire the
structure of WFs of the unperturbed Hamiltonian (\ref{1-ham1}). We
may expect that, in this case, the nodeless GS corresponds to a
liquid at any density, and the weak potential $U_{c}$ changes
insignificantly the solutions corresponding to Hamiltonian
(\ref{1-ham1}). As $U_{c}$ decreases to an infinitely small value,
the solutions should coincide with those for Hamiltonian
(\ref{1-ham1}).

We mention the idea to obtain the crystal nodeless GS from a liquid
GS as a result of the spontaneous breaking of the translational and
rotational invariances of the Hamiltonian (\ref{1-ham1}) due to the
introduction of an \textit{infinitely small} crystal potential
$U_{c}$ in (\ref{1-ham1}) \cite{guyer,bogquasi}. In our opinion,
such mechanism does not work. Indeed, it was supposed in
\cite{guyer,bogquasi} that the constant density (liquid GS) arises
due to the averaging over many configurations that differ from one
another only by translations. Therefore, the removal of the
translational invariance could lead to the crystal nodeless GS with
oscillating density. We think that this reasoning is not quite
correct. We note that, under a translation WF (\ref{1-new0000}),
(\ref{psip-2}) is only multiplied by a unit modulus constant.
Therefore, all states differing from one another only by
translations enter the statistical sum as \textit{one} state. In
other words, there is no averaging over such states. Moreover, the
constant density is characteristic of any WF (\ref{1-new0000}),
(\ref{psip-2}). The crystalline nodeless GS cannot be obtained as a
result of  the spontaneous choice of one of solutions
(\ref{1-new0000}), (\ref{psip-2}), or (\ref{1-5}), (\ref{1-2full}),
since the crystalline nodeless WF is simply absent among those WFs:
At any density, the nodeless WF corresponds to a liquid, as is shown
above.

Note that Hamiltonian (\ref{1-ham1}) is also invariant under the
group of permutations $S_{N}$. It is of interest that any symmetric
Bose solution is transformed by the identical (unit) representation
of the group $S_{N}$. Thus, the isotropic solution (\ref{1-5}),
(\ref{1-2full}) realizes the most symmetric state of an infinite
periodic system. This state is invariant with respect to all groups
of symmetries of the Hamiltonian: $O(3)$, $T(3)$, and $S_{N}$.
Apparently, for any BCs the ground state corresponds to the most
symmetric solution that is transformed by  the unit representations
of all groups of symmetries of the boundary-value problem
\cite{elliott}. Indeed, such a solution is non-degenerate and
changes most smoothly in space. Therefore, it should correspond to
the lowest energy.

We note also that for a finite ball the Hamiltonian and BCs are
characterized by the symmetry group $O(3)\times S_{N}$. Therefore,
GS should  be invariant with respect to these groups, which
corresponds to a liquid state. Thus, the lowest  state of a finite
crystal ball corresponds to WF with nodes for any parameters of the
system.

In the above reasoning, the degeneracy is related to the
noncommutativity of the operators $\hat{L}_{x}, \hat{L}_{y},
\hat{L}_{z}$. In work \cite{mtcrystal2}, the degeneracy of an
anisotropic state was proved for an infinite  Bose system with the
help of a general quantum-mechanical analysis. In this case, the
degeneracy is related to the noncommutativity of the operators
$\hat{\textbf{L}}$ and $\hat{\textbf{P}}$. In the two-dimensional
(2D) case, $\hat{\textbf{L}}$ has only the component $\hat{L}_{z}$.
In this case, only the proof from work \cite{mtcrystal2} is valid.

We now make some remarks about the method in use. It is impossible
to find the wave functions for a many-particle system of complicated
shape. Therefore, it is reasonable to take such BCs and such size of
the system for which the solution can be most easily found. The
properties of a macroscopic system should not depend on the shape of
boundaries and should not vary at an increase of the system sizes to
infinity (at a constant density). Therefore, we considered the
systems that  are the simplest for the symmetry analysis: the
infinite system under periodic BCs and a finite ball-shaped one
under zero BCs. For clarity, it would be useful to give the general
solution in the $\textbf{L}$-approach (besides the solutions
(\ref{1-new0000}), (\ref{psip-2}) in the $\textbf{P}$-approach).
However, such solutions have not been found in the available
literature. The degeneracy of an anisotropic state of the infinite
system can be seen without calculations: if the isotropic WF
$\Psi_{0}$ at some density transits spontaneously to an anisotropic
WF $\Psi_{0}^{\prime}$ with a separated vector $\textbf{k}$, then
exactly the same solutions $\Psi_{0}^{\prime}$ with a vector
$\textbf{k}^{\prime}$ ($|\textbf{k}^{\prime}|=|\textbf{k}|$) of all
possible directions must exist due to the isotropy of space. We
obtain the infinite-fold degeneracy. For a finite system, the
degeneracy is always finite \cite{gilbert}. This is due to BCs. For
example, for the periodic BCs, the momentum of the system of
quasiparticles can have only discrete values. Therefore, it can be
``turned'' only by certain discrete angles.

For a visual image we give several known anisotropic solutions for
an infinite periodic Bose system. The solutions for a liquid with
one and two phonons in the zero approximation reads
$\Psi_{\textbf{k}}(\textbf{r}_1,\ldots,\textbf{r}_N) =
\rho_{-\textbf{k}}\Psi_{0}$ \cite{bijl,bz1956,fey1972,yuv2} and
$\Psi_{\textbf{k}_{1}\textbf{k}_{2}}(\textbf{r}_1,\ldots,\textbf{r}_N)
= \rho_{-\textbf{k}_{1}}\rho_{-\textbf{k}_{2}}\Psi_{0}$
\cite{holes2020,fey1954,cbf-lee2}, respectively. These  solutions
are infinite-fold degenerate with respect to rotations. The crystal
solution for such a system was proposed in \cite{mtcrystal2}. It is
also infinite-fold degenerate. In all these cases the degeneracy is
caused by that the Hamiltonian $\hat{H}$ (\ref{1-ham1}) commutes
with the operators $\hat{\textbf{L}}$ and $\hat{\textbf{P}}$, which
do not commute with each other. Visually, the degeneracy is related
to the equivalence of all directions in space.

\subsection{Possible exact ansatz for the ground state of a Bose crystal}
Let us try to find WF of a zero-temperature crystal. We will call a
zero-temperature state of a crystal the ground state of a crystal.
It is obvious that the weakly excited states of the Bose system
cannot correspond to a crystal. In particular, the solution for the
liquid state with one phonon in the zero approximation reads
\cite{bijl,bz1956,fey1972,yuv2,mt2006}
 \begin{eqnarray}
    \Psi_{\textbf{k}}(\textbf{r}_1,\ldots,\textbf{r}_N) =
 \rho_{-\textbf{k}}\Psi_{0}, \quad \Psi_{0} =   e^{S_{0}}
       \label{1-14}\end{eqnarray}
(for periodic BCs). The solution for a two-phonon liquid state under
the same BCs in the zero approximation is
\cite{holes2020,fey1954,cbf-lee2}
 \begin{eqnarray}
\Psi_{\textbf{k}_{1}\textbf{k}_{2}}(\textbf{r}_1,\ldots
,\textbf{r}_N) = \rho_{-\textbf{k}_{1}}\rho_{-\textbf{k}_{2}}\Psi_0.
       \label{1-15}\end{eqnarray}
Therefore, it is clear that GS of a crystal should correspond to a
highly excited state of the Bose system. In addition, it is natural
to expect that the network of nodes of WF  does not change at the
displacement of the crystal by the lattice period.  The solution for
GS of a crystal can be easily guessed for a simple rectangular
lattice with $N\rightarrow\infty$ and the zero BCs ($\Psi=0$ at
$x=0; L_{x}$, $y=0; L_{y}$, $z=0; L_{z}$) \cite{zero-crystal}:
\begin{eqnarray}
 \Psi_{0}^{c}&=&e^{S_{0} +S_{c}} \prod\limits_{j=1}^{N}
 \{\sin{(k_{l_{x}}x_{j})}\sin{(k_{l_{y}}y_{j})} \sin{(k_{l_{z}}z_{j})}\}.
  \label{c-z}    \end{eqnarray}
Here, the product of sines directly sets the crystal lattice (we
suppose that faces coincide with lattice planes), $(k_{l_{x}},
k_{l_{y}}, k_{l_{z}})=(l_{x}\pi/L_{x}, l_{y}\pi/L_{y},
l_{z}\pi/L_{z})=(\pi/a_{x}, \pi/a_{y}, \pi/a_{z})$, $a_{x}, a_{y},
a_{z}$  are the periods of the lattice, $l_{x}, l_{y}, l_{z}$ are
integers, $L_{x}, L_{y}, L_{z}$ are sizes of the crystal; and
$S_{c}(\textbf{r}_{1}, \ldots, \textbf{r}_{N})$ is a correction
function. Function (\ref{c-z}) has a wave structure, but possesses a
lot of nodes, in contrast to (\ref{1-4}). Near any maximum $x_{0},$
the function $\sin(kx)$ can be represented as
$e^{-\alpha^{2}(x-x_{0})^{2}/2}$. This allows us to theoretically
get the fitting constant $\alpha$ with reasonable accuracy
\cite{zero-crystal}. Furthermore, if we use
$\varphi(\textbf{r})=-\alpha^{2}_{x} x^{2}/2-\alpha^{2}_{y}
y^{2}/2-\alpha^{2}_{z} z^{2}/2$ in (\ref{1-1}) instead of
$\varphi(\textbf{r})=-\alpha^{2} \textbf{r}^{2}/2$, then those
configurations, for which the atoms are located near the lattice
sites, are described by functions (\ref{c-z}) and (\ref{1-1})
equally (a more complicated ``nondiagonal'' function
$\varphi(\textbf{r})$ was considered in
\cite{koehler1966,nosanow1967}).  On the whole, WF (\ref{1-1}) can
be considered as a fairly good zero approximation. This property,
jointly with fitting parameters, enables one to explain with the
help of WF (\ref{1-1}) some experimental properties of crystals
\cite{nosanow1966,cazorla2008}. However, the general structure of
the wave function is represented by ansatz (\ref{1-1}) incorrectly.
In particular, ansatz (\ref{1-1}) loses the condensate of atoms
$\Psi_{c}(\textbf{r})\simeq \sin{(k_{l_{x}}x)}\sin{(k_{l_{y}}y)}
\sin{(k_{l_{z}}z)}$ which follows from WF  (\ref{c-z}). Moreover,
ansatz (\ref{1-1}) does not catch that GS of a crystal has to be
higher by energy than GS of a liquid (see Fig. 1).

Note that a possible ansatz for GS of a crystal with periodic BCs
was proposed in \cite{mtcrystal2}.

Comparing WF (\ref{c-z}) with the one-phonon (\ref{1-14})  and
two-phonon (\ref{1-15}) solutions for a liquid, we see that GS of a
crystal with the zero BCs corresponds to a liquid with $N$ identical
quasiparticles with quasimomentum $\textbf{k}_{\textbf{l}}$. That
is, GS of a crystal can be considered as a liquid with a condensate
of quasiparticles. In this case, \textit{namely the condensate of
quasiparticles creates a crystal lattice} in the medium. As was
mentioned above, GS (\ref{c-z}) contains also a condensate of
\textit{atoms} with quasimomentum $\textbf{k}_{\textbf{l}}$. Small
deviations from GS of a crystal correspond to a crystal with several
quasiparticles or defects. The temperature $T$ of a crystal can be
introduced in the ordinary way with the help of the partition
function, by connecting $T$ with quasiparticles.

We note that the solutions for a crystal that are characterized by a
condensate of atoms with quasimomentum $\textbf{k}\neq 0$  were
considered previously
\cite{gross1958,gross1960,coniglio1969,kirz,nep,mt2020}. However, it
was assumed in those works that, in addition to such ``coherent
crystal'' \cite{kirz,nep}, there exists the ``ordinary crystal''
with nodeless ground-state WF and without a condensate. But the
above analysis shows that such ``ordinary crystal'' is impossible,
at least for an infinite system. Moreover, the idea of that a
crystal is formed by a condensate of quasiparticles with
quasimomentum $2\pi/a_{x}$ [in one dimension (1D)] was advanced in
\cite{luca}. It is similar to the above conclusion, but the
quasiparticles here and in \cite{luca} are different. In function
(\ref{c-z}), the quasiparticles are introduced relative to the
genuine liquid GS of the system, whereas the quasiparticles in
\cite{luca} are considered relative to GS of a crystal.

The above analysis uses an anisotropy and is not suitable for a 1D
space. In the recent work \cite{mt1Dcrystal}, the \textit{exact}
solutions were found for a 1D system of point bosons with a small
value of $N$. In this case, the crystalline solution agrees with
formula (\ref{c-z}). We are not aware of other exact solutions for a
1D crystal which is not placed in a trap field. On the other hand,
for the 1D system of dipolar bosons, the crystal regime was
numerically found for the genuine nodeless GS (see the recent work
\cite{lode2020} and references therein). Thus, in one dimension, the
genuine GS can be either a liquid or a crystal, depending on the
nature of the interatomic interaction and on parameters of the
system.

We have noted above that the properties of a macroscopic system
should not depend on the shape of boundaries. It is a commonly
accepted assumption. However, it is not proved in the general case.
In our opinion, we cannot omit, in principle, the possibility of a
strong influence of boundaries as a topological effect. But we do
not know works,  where a similar effect is accurately found. The
available solutions show that the boundaries exert a negligible
influence on the bulk properties of a Bose liquid such as the energy
of GS (see the solutions for the periodic
\cite{bz1956,yuv1,girardeau1960,ll1963}, zero
\cite{gaudin1971,mt2015,mtmb2019}, and mixed \cite{bulatov1988} BCs)
and the dispersion law of quasiparticles
\cite{mtmb2019,mtsp2019,cazalilla2004}. In this case, the solutions
\cite{girardeau1960,ll1963,gaudin1971,mt2015,bulatov1988,mtsp2019}
are exact. Based on the above-executed analysis and those solutions,
we assume that the genuine GS of a Bose system corresponds to a
liquid at any shape of boundaries, any  density, and any
dimensionality of a system.

Monte Carlo solutions for Bose crystals are discussed in Appendix 3.
The nature of GS of a Bose system can be clarified by the
multiconfiguration time-dependent Hartree method \cite{MCH} that
allows one to find a solution with good accuracy for a 1D system of
$N\lsim 10$ bosons. Apparently, the last modifications of this
method \cite{meyer2019} enable one to study even 2D systems of
$N\lsim 10$ bosons. This method is also suitable for the study of
the transition from the crystalline genuine GS to the liquid one, as
the bare crystal potential $U_{c}$ decreases, for  1D and 2D systems
of $N\lsim 10$ particles. Such results would be  valuable.

As is seen, the properties of Bose crystals are, apparently, much
more complex and interesting, than it follows from the ``naive'' WF
(\ref{1-1}).

 \section{Physical consequences}
On the basis of the above analysis, we assume that, for any BCs, the
inequality
\begin{eqnarray}
   E_{0}^{c}(\rho, N) >  E_{0}^{l}(\rho, N)
       \label{e00}\end{eqnarray}
holds. Here, $E_{0}^{c}$ and $E_{0}^{l}$ are the energies of GS of a
Bose crystal and a Bose liquid, respectively, $N$ is the number of
atoms, and $\rho=mn$ is the density. In (\ref{e00}), $E_{0}^{c}$ and
$E_{0}^{l}$ are compared at the same $\rho$. However, the phase
transitions occur in experiments at the same pressure $P$. Here, two
cases are possible:
\begin{eqnarray}
   E_{0}^{c}(P,N) >  E_{0}^{l}(P,N)
       \label{e01}\end{eqnarray}
or
\begin{eqnarray}
   E_{0}^{c}(P,N) <  E_{0}^{l}(P,N).
       \label{e02}\end{eqnarray}
For $^4$He, inequality (\ref{e01}) is satisfied (at the pressure of
crystallization $P\approx 25\,atm$, see Appendix 3). The liquid
satisfying condition (\ref{e01}) must be stable against
crystallization, at low $P$ and $T$. If (\ref{e02}) is satisfied,
the liquid corresponds to a metastable state, but the duration of
the transition into the stable crystalline state may be long.

Inequality (\ref{e00}) testifies to the existence of a large number
of quantum states corresponding to a liquid and possessing the
energies less than the GS energy  of a crystal. We will call such
states ``under-crystal liquid'' (``underliquid'' for short). Since
this region of states is large, one can expect that at least part of
it is observable. It is also clear that, at sufficiently low
temperatures, the underliquid has to be superfluid. The creation of
such superfluids will mean that, in addition to the vessels with He
II, physical laboratories will possess the vessels with other
superfluids. Let us try to ascertain how the underliquid can be
produced.

For all known liquids, except for $^4$He, the $(P,T)$ phase diagram
is separated into the regions corresponding to a gas, a liquid, and
a  crystal and has the triple point (see Fig. 2). The $(P,T)$
diagram of $^4$He has no triple point:  the gas contacts only with
the liquid. Each of the transitions (gas--liquid, liquid--crystal,
and gas--crystal) is operated by three equations describing the
equilibrium between phase 1 and phase 2 \cite{gibbs}:
$P_{1}=P_{2}\equiv P$, $T_{1}=T_{2}\equiv T$, and
\begin{eqnarray}
P(v_{1}-v_{2})+T[s_{2}(P,T)-s_{1}(P,T)]=E_{2}(P,T)-E_{1}(P,T),
       \label{ee-clc}\end{eqnarray}
where  $E_{j}$ is the internal energy per atom for the system
staying in the $j$-th phase, $v_{j}$ and $s_{j}$
are the volume and entropy (per atom) of the $j$-th phase. Equation (\ref{ee-clc})
is equivalent to the equality of the chemical potentials of phases 1 and 2:
$\mu_{1}(P,T)=\mu_{2}(P,T)$.

\begin{figure}[ht]
\centerline{\includegraphics[width=85mm]{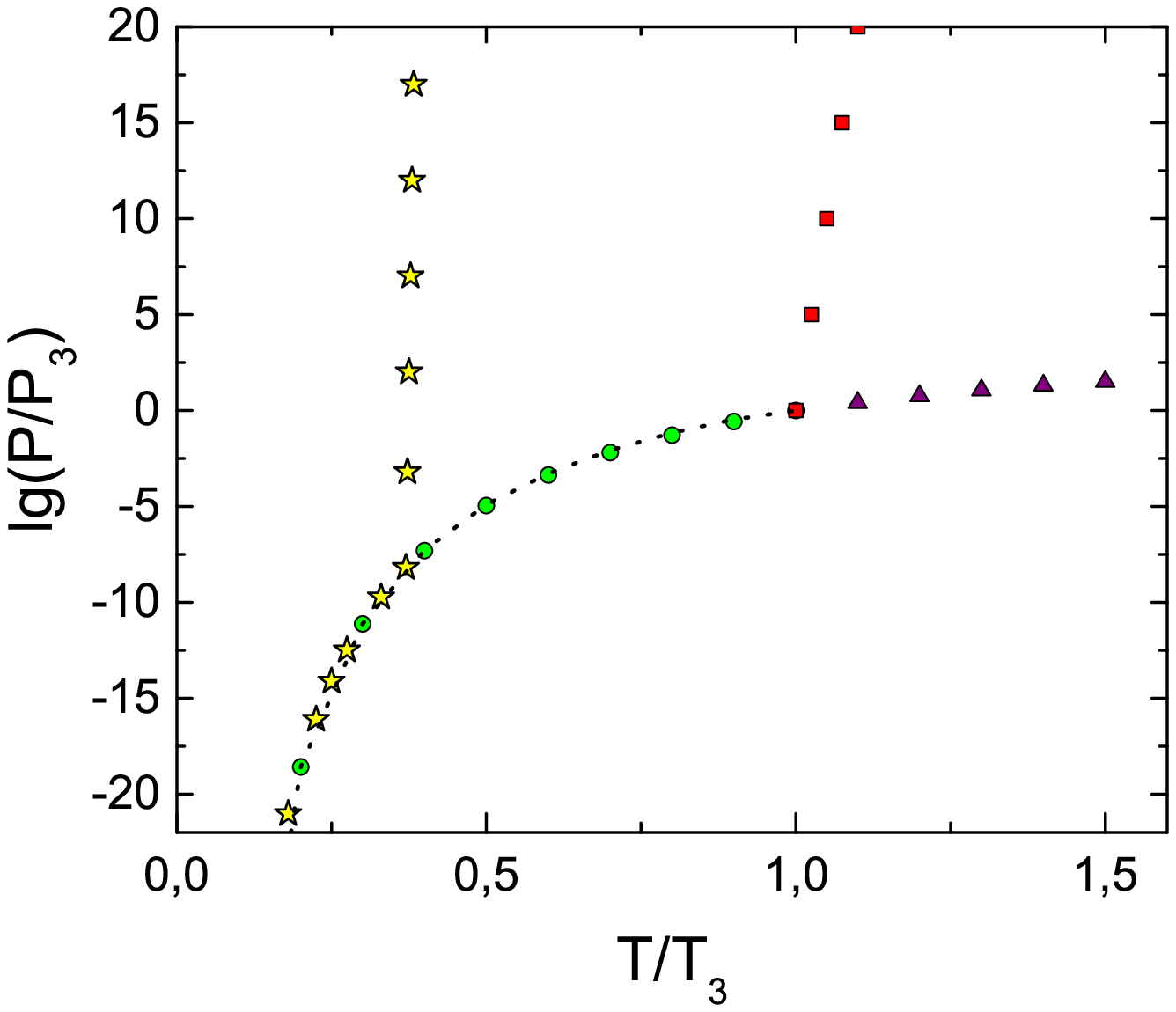} } \caption{
[Color online]  $(P,T)$ phase diagram for inert elements (H$_{2}$,
Ne, Ar, etc., except for $^{4}$He) with the assumed region of the
under-crystal liquid (bounded by stars $\star\star\star$); $\lg
\equiv\log_{10}$. Lines of the transitions gas--crystal
($\circ\circ\circ$, Eq. (\ref{PT1}) with $\xi= 9.69$), gas--liquid
(${\sss\blacktriangle\blacktriangle\blacktriangle}$), and
liquid--crystal (${\sss\blacksquare\blacksquare\blacksquare}$) are
shown. $P_{3}$ and $T_{3}$ are the pressure and temperature at the
triple point. Dotted line (Eq. (\ref{PT2}) with $\xi= 9.69$, $f=7$)
marks the continuation of the line gas--liquid to the region of low
$T$. This curve lies somewhat higher than the curve
$\circ\circ\circ$, but the difference is visually indistinguishable
(maximum distance between the curves along the vertical is equal to
$0.13$ and corresponds to $T/T_{3}\approx 0.8$). The curves
${\sss\blacktriangle\blacktriangle\blacktriangle}$,
${\sss\blacksquare\blacksquare\blacksquare},$ and the right vertical
boundary of the region of the under-crystal liquid are drawn by eye.
 \label{fig2}}
\end{figure}

The possible position of the underliquid region  on the $(P,T)$
diagram is shown by stars in Fig. 2. The upper and lower boundaries
of the underliquid region are set by condition (\ref{ee-clc}). The
lower boundary corresponds to the transition underliquid--gas. The
possible upper boundary corresponds to the transition
underliquid--crystal. In the limits of these boundaries, the liquid
can be stable or metastable, as was noted above. The right boundary
$P_{ul-c}(T)$ is shown in Fig. 2 approximately. It corresponds not
to a phase transition, but to the condition $E^{l}(P,T)=
E_{0}^{c}(P)$ (that is possible, if (\ref{e01}) is satisfied). The
equation for the lower boundary will be obtained in what follows. It
is easy to estimate the location of the upper boundary
$P_{ul-c}(T)$. At $T=0,$ relation (\ref{ee-clc}) yields
\begin{eqnarray}
P_{ul-c}(T=0)=\frac{E_{0}^{c}(P)-E_{0}^{l}(P)}{v_{l}-v_{c}}.
       \label{pp-ul-c}\end{eqnarray}
For the realistic values $v_{l}-v_{c}=0.1 v_{l}= 0.1(3.6{\mbox
\AA})^3$ and $E_{0}^{c}(P)-E_{0}^{l}(P)=10K k_{B},$ we find
$P_{ul-c}(T=0)\approx 300\,atm$ (here, $k_{B}$ is the Boltzmann
constant). The upper boundary exists, if $P_{ul-c}(T=0)> 0$. At
$P_{ul-c}(T=0)< 0$ the upper boundary is absent, which corresponds
to Fig. 2. This case is considered below in more details.

We now estimate the temperature for the right boundary ($E^{l}(P,T)=
E_{0}^{c}(P)$) at $P\approx 0$. According to  \cite{guyer},  the
relation $k_{B}T_{3}\approx 2\varepsilon/3$ holds for the inert
elements (here, $\varepsilon$ is the energy from the Lennard-Jones
potential). Assume that $E^{l}_{0}(P=0)- E_{0}^{c}(P=0)\sim
-0.1\varepsilon \sim -0.15k_{B}T_{3}$, similarly to $^{4}$He. At low
$T,$ we may consider only the phonon contribution to the energy.
Then $E^{l}(T)= E_{0}^{l}+\frac{\pi^{2}}{30}\left (
\frac{k_{B}T}{\hbar c_{s}}\right )^{3}\frac{k_{B}T}{n}$ \cite{khal},
where $c_{s}$  is the first sound velocity. The relations
$E^{l}(P=0,T)= E_{0}^{c}(P=0)$ and $E^{c}_{0}(P=0)-
E_{0}^{l}(P=0)\simeq 0.15k_{B}T_{3}$ yield
\begin{eqnarray}
\frac{T^{4}}{T_{3}^{4}}\simeq\frac{30\cdot 0.15n}{\pi^{2}} \left (
\frac{\hbar c_{s}}{k_{B}T_{3}}\right )^{3}.
       \label{Tr}\end{eqnarray}
Using the parameters of the triple point for neon ($T_{3}=24.55\,K$
 \cite{pollack1964,glazov,pavese},  $c_{s}=628\,m/s$,
$\rho=1.24\,g/cm^3$ \cite{pollack1964,naugle}), we get  $T\approx
0.6T_{3}$. For liquid argon at the triple point, we have
$T_{3}=83.81\,K$
 \cite{pollack1964,glazov,pavese},  $c_{s}=870\,m/s$,
$\rho=1.38\,g/cm^3$ \cite{itterbeek}. In this case, we obtain from
(\ref{Tr}) $T\approx 0.26T_{3}$. We expect that these estimates of
$T$ are valid by the order of magnitude.

The basic question is: How can we ``get to'' the region of
underliquid in experiments? (The underliquid state of $^4$He has
already been obtained: as it is easy to guess, this is He II.) On
top and to the right from the region of underliquid, the crystalline
states are placed. The region of underliquid corresponds to very low
temperatures: $T\lsim 0.5T_{3}$. The crystalline states at such $T$
were experimentally studied for many substances, but the underliquid
states were not found. According to (\ref{e01}) and (\ref{e02}), the
crystal with $T$ and $P$ from the region of underliquid should be
metastable or stable, respectively. In the metastable state, such
crystals live, apparently, very long (otherwise, the underliquid
would be found experimentally long ago). Therefore, we believe that
it is impossible to produce an underliquid from a crystal (by
decreasing $T$ or $P$).

The underliquid can be, apparently, obtained by strong supercooling
of a liquid whose initial temperature is higher than the melting
one. To avoid the crystallization, it is necessary to purify a
liquid from impurities and to use a vessel with smooth walls (or to
cover the walls with a special coating, see below).  A shortcoming
of the method consists in the necessity of a strong supercooling,
which requires the high degree of smoothness of walls and of purity
of a liquid.

It may be easier to get in the underliquid region by isothermal
compression of a gas at $T\ll T_{3}$. According to Fig. 2, at the
isothermal increase in the pressure of a gas with $T\ll T_{3},$ we
cross firstly the curve gas--crystal and then the curve gas--liquid.
Therefore, the gas must turn into a crystal (not in the
underliquid), which corresponds to experiments. Nevertheless, we
will show below that the underliquid can be obtained in such a way.
For this purpose, it is necessary to perform the transition at the
temperature $T\ll T_{3}$ and to create the conditions preventing the
crystallization.

To substantiate this point, we consider the transitions gas--crystal
(g-c) and gas--liquid (g-l) in more details. First, one needs to get
the dependences $P(T)$ setting the curves g-c and g-l. As is known,
along the line $P(T)$ of the phase transition the
Clapeyron--Clausius relation should hold:
\begin{eqnarray}
\frac{d P}{d T}= \frac{s_{1}(P,T)-s_{2}(P,T)}{v_{1}-v_{2}}.
       \label{cc}\end{eqnarray}
Let index 1 correspond to a gas, and index 2 to a liquid or a
crystal. The data on the pressure of saturated vapors for He II
\cite{eselson1978} show that, at $T\lsim T_{\lambda},$ the
temperature of a vapor is much larger than the temperature of the
Bose condensation. Therefore, the vapor can be considered as an
ideal gas. Assume that the vapors of other inert elements at $T\leq
T_{3}$ can also be considered as an ideal gases. The entropy of a
one-atom ideal gas consisting of atoms with zero spin and zero
orbital moment (all inert elements, except for $H_{2}$) is known
\cite{land5,huang}:
\begin{eqnarray}
s_{g}=\frac{5}{2}k_{B}+k_{B}\ln{ \left
[\frac{(k_{B}T)^{5/2}}{P}\left (\frac{m}{2\pi \hbar^{2}}\right
)^{3/2} \right ]}.
       \label{sg}\end{eqnarray}
Note that the first term in this formula is given in the literature
with different coefficients: $3/2$ \cite{huang}  and $5/2$
\cite{land5}. We did not study the reason for this difference and
will use $5/2$ (for the estimates below, the  difference between
$5/2$ and $3/2$ is insignificant).

The entropy $s_{2}$ of a liquid or crystal at $T\leq T_{3}$ is
determined mainly by the phonon contribution ($\sim T^{3}$), which
is much less than the entropy $s_{1}=s_{g}$ of a saturated vapor (we
remark that the Debye temperature for inert elements is comparable
with $T_{3}$). The entropy of a Bose liquid at $T\rightarrow 0$
reads \cite{khal}
\begin{eqnarray}
s_{l}=\frac{2\pi^{2}}{45}\frac{k_{B}}{n} \left ( \frac{k_{B}T}{
\hbar c_{s}}\right )^{3}.
       \label{sl}\end{eqnarray}
For  $^{4}He$ atoms at $T=1\,K$ and at the pressure of saturated
vapors $P\approx 1.6\cdot 10^{-4}\,atm$ \cite{eselson1978}, we get
$s_{g}/s_{l}\simeq 3000$. For neon at the triple point, we have
$P_{3}=0.427\,atm$ \cite{glazov} and $s_{g}/s_{l}\approx 7.7$. These
estimates indicate that, in the region of temperatures from $0\,K$
to $T_{3},$ the relation $s_{l}, s_{c} \ll s_{g}$ holds. In this
case, $v_{l}, v_{c} \ll v_{g}$. Therefore, in the zero
approximation, the curves gas-crystal and gas-liquid at $0\leq T\leq
T_{3}$ are given by the formula
\begin{eqnarray}
\frac{d P}{d T}= \frac{s_{g}}{v_{g}}=\frac{P s_{g}}{k_{B}T}.
       \label{cc0}\end{eqnarray}
Denote $\tilde{T}=T/T_{3}$, $\tilde{P}=P/P_{3}$. Then
formula (\ref{sg}) can be written as
\begin{eqnarray}
s_{g}/k_{B}=\frac{5}{2}\ln{\tilde{T}}-\ln{\tilde{P}}+s_{3},
       \label{sg2}\end{eqnarray}
where $s_{3}$ is the value of $s_{g}/k_{B}$ at the triple point.
Equation (\ref{cc0}) takes the form
\begin{eqnarray}
\frac{d \tilde{P}}{d \tilde{T}}= \frac{\tilde{P}}{\tilde{T}}\left
(\frac{5}{2}\ln{\tilde{T}}-\ln{\tilde{P}}+s_{3}\right ).
       \label{cc01}\end{eqnarray}
Now, denote $y=\ln{\tilde{P}}$ and $x=\ln{\tilde{T}}$. Then Eq.
(\ref{cc01}) becomes
\begin{eqnarray}
\frac{d y}{d x}= \frac{5}{2}x-y+s_{3}.
       \label{cc02}\end{eqnarray}
We need to find a solution satisfying the boundary condition $y=x=0$ (for the
triple point). The solution can be sought as a series
$y=a_{1}x+a_{2}x^{2}+\ldots +a_{j}x^{j}+\ldots$. After the simple transformations,
we get
\begin{eqnarray}
y= \xi + \frac{5x}{2}-\xi e^{-x}, \quad \xi=s_{3}-\frac{5}{2}.
       \label{y1}\end{eqnarray}
From (\ref{y1}) we obtain $P(T)$ for a saturated vapor at $0< T \leq
T_{3}$:
\begin{eqnarray}
\tilde{P}=e^{\xi}\tilde{T}^{5/2}e^{-\xi/\tilde{T}}.
       \label{PT1}\end{eqnarray}
This is a solution in the zero approximation. It holds for both
curves g-l and g-c. We do not know whether this solution was
obtained previously.

$^{4}He$ has no triple point. If  we set formally $T_{3}=1\,K$ for
$^{4}He$, then Eq. (\ref{PT1}) agrees very well with the
experimental pressure $P(T)$ of saturated vapors of $^{4}He$ at
$T\leq 1\,K$ \cite{eselson1978}. If we take $T_{3}=3\,K$, then Eq.
(\ref{PT1}) describes experiments only qualitatively (perhaps
because formula (\ref{cc0}) becomes a poor approximation for
(\ref{cc})). For Ne, Ar, Kr, and Xe, the dependence $P(T)$ for the
sublimation curve was measured for temperatures $T\simeq (2/3 \div
1)T_{3}$ \cite{pavese,rab}. In particular, the experimental
dependence $P(T)$ for neon at $T=16-24\,K$ is described by the
fitting formula $\lg{\tilde{P}}\approx
3.2-21.39\lg{\tilde{T}}+5.4\tilde{T}-8.6/\tilde{T}$ \cite{rab}. The
approximate solution (\ref{PT1}) with $\xi$ for neon ($\xi= 9.69$)
gives the values of $\lg{\tilde{P}}$ less by about $10\%$.

Solution (\ref{PT1}) was found by neglecting the corrections $s_{2}$
and $v_{2}$ in (\ref{cc}). At $\tilde{T}\ll 1$ these corrections are
negligible. They increase with $\tilde{T}$, but remain small even at
$\tilde{T}=1$. In order to estimate the influence of corrections on
the solution, we take the entropy
$s_{2}/k_{B}=4\tilde{T}^{3}-(f+1)\tilde{T}^{f}$ with $f>3$  into
account in (\ref{cc}). Here, the first term describes $s_{2}/k_{B}$
proper (for comparison, $s_{l}/k_{B}=1.58\tilde{T}^{3}$ for neon at
$\tilde{T}\ll 1$; while estimating $s_{l},$ we take $c_{s}(T\leq
T_{3})=c_{s}(T_{3})$ and $\rho(T\leq T_{3})=\rho(T_{3})$). The
second term effectively describes  $v_{2}$ from the denominator. In
this case, we get the solution
\begin{eqnarray}
y= \xi + \frac{5x}{2}-\xi e^{-x}+e^{3x}-e^{fx},
       \label{y2}\end{eqnarray}
\begin{eqnarray}
\tilde{P}=e^{\xi}\tilde{T}^{5/2}e^{-\xi/\tilde{T}}e^{\tilde{T}^{3}-\tilde{T}^{f}}.
       \label{PT2}\end{eqnarray}
In Fig. 2, this solution is shown as the curve g-l and solution
(\ref{PT1}) as the curve g-c.  Of course, such correspondence
between the formulae and the curves is only qualitative.
For Fig. 2 we use parameter $\xi= 9.69$ corresponding to neon. In
this case, the parameter $f=7$ is chosen so that curve (\ref{PT2})
lies above curve (\ref{PT1}), and the slope of curve (\ref{PT2}) at
$T\rightarrow T_{3}$ is less than that of curve (\ref{PT1}). As a
result, curves (\ref{PT1}) and (\ref{PT2}) are similar to
experimental curves g-c and g-l, respectively. The exact curves g-l
and g-c can significantly differ from those presented in Fig. 2,
because the corrections $s_{2}$ and $v_{2}$ were taken into account
in a rough model form. However, this analysis is sufficient to show
that the solutions of such type correctly describe experimental
curves g-l and g-c. The second important conclusion is that though
the corrections $s_{2}$ and $v_{2}$ separate the curves g-l and g-c,
\textit{these curves should be close.} For example, at $\tilde{T}=
0.1$ we have $P_{gl}/P_{gc}=e^{\tilde{T}^{3}-\tilde{T}^{f}}\approx
1.001$, according to relations (\ref{PT1}) and (\ref{PT2}) with
$f=7$.

It is significant that, for inert elements, the slopes of the
experimental $P(T)$ curves g-l (at $T> T_{3}$) and g-c (at $T<
T_{3}$) near the triple point are very close \cite{glazov}. This
agrees with our conclusion that these curves should be close at
$T\leq T_{3}$.

If the ratio $P_{gl}/P_{gc}=\zeta$ is close to 1, the phase
transitions gas--crystal and gas--liquid are ``switched-on''  almost
simultaneously. At the compression, the system transits in a liquid
or a crystal depending on  that which nuclei are generated faster:
microdrops or microcrystals. In Appendix 4, where the formation of
nuclei is considered, we will show that it is necessary to increase
the pressure of a gas up to $P\approx P_{gc}
\zeta^{\frac{1+\phi}{\phi}}$ in order that the microdrops are
generated faster, than microcrystals ($\phi$ depends on the
substance; the characteristic value is $\phi\simeq 0.1$). If we use
a vessel with smooth walls whose microstructure differs
significantly from that of crystal nuclei, and if a gas is purified
from impurities, then the formation of crystal nuclei should be
suppressed, though the curve $P_{gc}(T)$ lies below  the curve
$P_{gl}(T)$ (see Appendix 4). Let the gas be compressed at the
pressure $P\approx P_{gc} \zeta^{\frac{1+\phi}{\phi}}\sim P_{gc}
\zeta^{11}\sim 1.01 P_{gc}$ (for $\zeta=1.001$, according to the
above estimate). Then one can expect that the gas will be condensed
into a liquid. If the energy of this liquid
$E^{l}(T,P)<E_{0}^{c}(P)$, then such underliquid will not
crystallize. Of course, our estimates are crude, and exact formulae
can give a much larger ratio $P/ P_{gc}$. However, we expect that
$\zeta^{\frac{1+\phi}{\phi}}\lsim 2$, i.e., the pressure $P$ should
be increased by at most several times as compared with $P_{gc}$, in
order that the spontaneous (bulk or surface) condensation of a gas
into a liquid to begin.

According to the analysis in Appendix 4, in order to prevent the
crystallization of a gas and to ``switch-on''  the bulk spontaneous
mechanism of formation of nuclei, one needs to purify a gas from
suspended solid impurities and to prevent the formation of
crystalline nuclei on the walls. To achieve the latter, one can take
a vessel with smooth walls (though, it is impossible to obtain an
ideally smooth walls), and the molecules of walls should weakly
interact with the molecules of a gas  (or the crystalline ordering
of walls should significantly differ from that of crystal nuclei
forming from a gas). In addition, the molecular dynamics simulations
show that the crystallization of a liquid on walls is suppressed, if
the walls are covered with a solid amorphous layer whose structure
is similar to that of a liquid \cite{espinosa2019} (see also
\cite{sosso2016}, Sect. 2.4.2). We propose one more method: One can
cover the walls from inside by a microscopically thin film of He II,
then the surface of walls should be liquid and smooth. In this case,
the formation of crystal nuclei on the walls would become difficult.
Moreover, the interaction of helium atoms with molecules of the
majority of gases is weak, which must prevent the adsorption of
molecules of a gas on the walls and the formation of surface nuclei.

It is noted in books \cite{strickland,frenkel} that, at the
compression of a gas at  a temperature $T<T_{3}$, the metastable
liquid is sometimes formed and then crystallizes. These properties
are evidence of the validity of the inequality $E_{0}^{c}(P) <
E_{0}^{l}(P)$ (\ref{e02}). However, our analysis shows that, for
some substances, the inequality $E_{0}^{c}(P) > E_{0}^{l}(P)$
(\ref{e01}) should hold. In this case, the liquid formed at the
compression of a gas should be stable and should not crystallize.

Interestingly, the transition  crystal--underliquid can occur at a
negative pressure. By (\ref{pp-ul-c}), we have $P_{ul-c}(T=0)< 0$
for $v_{l}-v_{c}> 0$, $E_{0}^{c}(P)-E_{0}^{l}(P)< 0$ or for
$v_{l}-v_{c}<  0$, $E_{0}^{c}(P)-E_{0}^{l}(P)>  0$. We may expect
that $P_{ul-c}(T=0)\sim -(100\div 1000)\,atm$. In this case, the
state of underliquid can apparently be obtained by creating a
negative pressure in a crystal. The idea of the creation of a liquid
from a crystal by applying a negative pressure was advanced by J.
Frenkel \cite{frenkel,frenkel1935}.

The above analysis shows that the form of the $(P,T)$-diagram at low
$P$ and $T$ should depend on how we got in this region. If we have
got into it from  above (by cooling a crystal), we obtain an
ordinary diagram with the triple point. The result will be the same,
if we will try to get into the region of underliquid from the bottom
(from the region of a gas) or from the right (from the region of a
liquid) in the absence of the conditions preventing the formation of
crystal nuclei. But if we try to enter from the bottom (or from the
right) in the presence of such conditions, then we should obtain the
state of underliquid. In other words, the phase diagram in the
region to the left from the melting curve and above the sublimation
curve should have two levels (or two ``layers''): liquid-type and
crystal-type ones. Such liquid states were obtained previously by
supercooling a liquid. In this case, it was considered in the
literature that the liquid state at $T=0$ is impossible. Therefore,
the supercooling of a liquid down to $T\approx 0$ and the obtaining
of a stable liquid by the compression of a gas at $T\approx 0$ were
also considered impossible. However, both are possible, since a
liquid can have the zero temperature, as was shown in Sect. 2.

Our reasoning is general and should be suitable for any inert
element (H$_{2}$, Ne, Ar, etc., see review \cite{aziz1984}), except
for helium. \textit{We propose to carry out three following
experiments}. (i) To supercool isobarically liquid H$_{2}$, Ne, and
Ar down to temperatures that are by several times less than the
temperature of the Bose condensation of the ideal gas
$T_{c}=\frac{3.31}{(2s+1)^{2/3}}\frac{\hbar^{2}}{k_{B}m}n^{2/3}$
\cite{land5} (here, $s$ is the spin of a particle; for liquid inert
elements, except for hydrogen and helium, $T_{c}<1\,K$). In this
case, we should obtain a superfluid underliquid. We note that, at a
strong overcooling, the energy barrier of the nucleation for some
liquids, apparently, disappears (see \cite{sosso2016}, Sect. 1.1.4
and 2.2.2). Such liquids can easily crystallize spontaneously. One
needs to pass through this temperature region and to get lower
$T\lsim T_{c}$ at which the microcrystals should become unstable
(according to the above analysis). (ii) To compress isothermally a
dilute gas (H$_{2}$, Ne, Ar) at $T\sim 0.1T_{3}$ up to a pressure
that exceeds by several times the saturated vapor pressure for the
same $T$. The underliquid should also be created. In both
experiments, it is necessary to create the conditions hampering the
crystallization (see the discussion above and Appendix 4), and
condition (\ref{e01}) must be satisfied for the stability of a
liquid. In the second experiment, a less supersaturation is
required. Therefore, the requirements to the purification and to
walls can be apparently less strict. (iii) To create a negative
pressure $P_{ul-c}\lsim -1000\,atm$ in a crystalline inert element
at $T\lsim 0.1T_{3}$. One can expect that some of such crystals will
melt into an underliquid.

If the underliquid is metastable (inequality (\ref{e02})) and has a
small life-time, then such  underliquid state can be unobservable.
However, $^4$He is stable (inequality (\ref{e01})). Therefore, it is
natural to expect that, among inert elements, there are several
other ones with the stable underliquid state. In view of this, it is
desirable to execute the experiments with all inert elements (except
for $^4$He).

The inert elements were investigated mainly at $T\gsim T_{3}$. The
number of experiments at $T\ll T_{3}$ is much less. In the last ones
the crystals, being in equilibrium with their vapors, were studied
at $T\geq 1\,K$ \cite{pavese,rab}.  \textit{We assume that the
underliquid state was not obtained earlier because the conditions
hampering the crystallization were not created.} Therefore, a gas or
liquid turned into a crystal, rather than into an underliquid.
Moreover, the underliquid state was considered impossible and was
not sought.

Why do the liquids in the Nature crystallize at the cooling, though
the crystal corresponds to a highly excited state of a system?
Mathematically, this is related to the fulfillment of condition
(\ref{ee-clc}) for the liquid--crystal transition. The possible
physical explanation is as follows: at the cooling of a liquid down
to some temperature, the microcrystals arising as fluctuations
become stable. And the visual reason is that the system falls into
the local energy minimum corresponding to a crystal (see Fig. 1). As
a result, the liquid crystallizes, and we obtain a crystal with some
number of quasiparticles. In this case, the cooling of the crystal
means a decrease in the number of quasiparticles introduced relative
to GS of a crystal.

We note also that, at the strong supercooling, the viscosity of some
liquids increases sharply, and they transit into a glass-like state.
However, the atoms of inert gases are spherically symmetric (except
for hydrogen which forms molecules $H_{2}$) and, therefore, should
not turn into a glass at the supercooling. We may expect that, at
$T\lsim 0.1T_{c}$, liquid inert elements have to be similar to He
II, i.e., they should contain a condensate of atoms with zero
momentum and should be superfluid.

Undoubtedly, inequality  (\ref{e00}) should be correct. Therefore,
the region of underliquid must exist, and the task is to enter this
region in experiments.

In Appendix 3, we also consider the third principle of
thermodynamics and properties of $^4$He.

 \section{Conclusions}
Our analysis shows that the genuine ground state of a system of
spinless bosons should correspond to a liquid or gas,  at any
density. \footnotemark\ \footnotetext{Such idea was proposed
previously in \cite{zero-crystal}, but that work of ours is immature
and contains something very similar to errors; see, in particular,
the Introduction in \cite{mtmb2019}} We have proved this for an
infinite 3D system and a finite ball-shaped 3D one. It is natural to
expect that this assertion is valid for finite systems of  any
shape. In this case, the lowest states of a liquid and a crystal
must satisfy the inequality $E_{0}^{c}(P)
> E_{0}^{l}(P)$ (\ref{e01}) or $E_{0}^{c}(P) < E_{0}^{l}(P)$
(\ref{e02}). If inequality (\ref{e02}) holds, the stable state of
the system at $T\rightarrow 0$ is a crystal, that corresponds to the
available experimental data. However, we expect that relation
(\ref{e01}) holds for the majority of inert elements (in particular,
it holds for $^4$He). The underliquid state, that does not
crystallize at the cooling and is superfluid at very low
temperatures, should exist for such substances. This is our main
experimental prediction. We assume that the underliquid can be
created in experiments by compressing a gas at a low temperature or
by strong supercooling an ordinary liquid (in both cases, it is
necessary to create the conditions preventing the crystallization).

According to the above analysis, a Bose crystal is a standing wave
in the probability field. Most likely, this property is a general
principle valid not only for Bose systems. Therefore, it is possible
that the underliquid state  and the superfluidity are inherent not
only in inert elements. Such properties can be inherent in all
substances that form molecular crystals: inert elements, water,
methane, etc. Although it may seem implausible now.

If inequality (\ref{e00}) is true under any boundary conditions, it
will change our comprehension of the nature of crystals and lead to
the discovery of new physical phenomena. For example, the space
apparatus ``New Horizons'' found in 2015 that the Sputnik Planitia
surface on Pluto consists of solid nitrogen and is similar to a
mosaic made of hexagons and pentagons. This valley has no craters,
though they are present on the remaining Pluto's surface. It is
possible that a liquid water ocean exists under the surface
\cite{pluto1,pluto2,pluto3}. That is surprising because the Pluto's
surface temperature is about $40\,K$. However, we have established
above that the liquids of low viscosity can exist even at $T= 0\,K$.
This can help one to understand some anomalous properties of cosmic
objects.

We hope that the above-proposed experiments to create the
underliquid state will be carried out.

The present work is partially supported by the National Academy of
Sciences of Ukraine (project No.~0121U109612).



\section{Appendix 1. Wave functions of a many-particle Bose system in the momentum approach}
We now determine the general form of WF of the ground state and any
excited state of a periodic Bose system on the basis of the analysis
in \cite{yuv1}. Since work \cite{yuv1} is not widely available, we
give below the derivation of these formulae.  Consider the functions
\begin{equation}
                    \psi^{f}_{0}=c_{00},
\label{psi-1} \end{equation}
\begin{equation}
                    \psi^{f}_{\textbf{k}_{1}} =  c_{11}\rho_{-\textbf{k}_{1}},
\label{psi-2} \end{equation}
\begin{equation}
\psi^{f}_{\textbf{k}_{1}\textbf{k}_{2}} = c_{22}\left
(\rho_{-\textbf{k}_{1}}\rho_{-\textbf{k}_{2}}-\frac{\rho_{-\textbf{k}_{1}-\textbf{k}_{2}}}{\sqrt{N}}
\right ), \label{psi-3} \end{equation}
\begin{eqnarray}&&
\psi^{f}_{\textbf{k}_{1}\textbf{k}_{2}\textbf{k}_{3}} = c_{33}\left
[\rho_{-\textbf{k}_{1}}\rho_{-\textbf{k}_{2}}\rho_{-\textbf{k}_{3}}-\frac{1}{\sqrt{N}}\left
(
\rho_{-\textbf{k}_{1}}\rho_{-\textbf{k}_{2}-\textbf{k}_{3}}+\rho_{-\textbf{k}_{2}}\rho_{-\textbf{k}_{1}-\textbf{k}_{3}}
+\rho_{-\textbf{k}_{3}}\rho_{-\textbf{k}_{1}-\textbf{k}_{2}}\right
)+\right. \nonumber
\\ &&+\left.\frac{2}{N}\rho_{-\textbf{k}_{1}-\textbf{k}_{2}-\textbf{k}_{3}}
\right ], \label{psi-4} \end{eqnarray}
\begin{equation}
\ldots, \label{psi-5} \end{equation}
\begin{eqnarray}
\psi^{f}_{\textbf{k}_{1}\ldots\textbf{k}_{N}} &=&
c_{N1}\rho_{-\textbf{k}_{1}}\cdots\rho_{-\textbf{k}_{N}}+c_{N2}\sum\limits_{P(\textbf{k}_{j})}
\rho_{-\textbf{k}_{1}}\cdots\rho_{-\textbf{k}_{N-2}}\rho_{-\textbf{k}_{N-1}-\textbf{k}_{N}}+\nonumber
\\ &+& \ldots+c_{NN}\rho_{-\textbf{k}_{1}-\ldots-\textbf{k}_{N}},
\label{psi-6} \end{eqnarray} where $c_{ij}$ are constants, and
$\sum_{P(\textbf{k}_{j})}$ is the sum over all permutations of the
vectors $\textbf{k}_{j}$. These are the wave functions of a periodic
system of $N$ \textit{free} spinless bosons. Here, $\psi^{f}_{0}$
corresponds to the ground state; $\psi^{f}_{\textbf{k}_{1}}$
describes the state, where one boson has a momentum
$\hbar\textbf{k}_{1},$ and $N-1$ bosons have the momentum $0$; and
so on; $\psi^{f}_{\textbf{k}_{1}\ldots\textbf{k}_{N}}$ describes the
state in which each of the bosons has some nonzero momentum
$\hbar\textbf{k}_{j}$. These functions are solutions of the
Schr\"{o}dinger equation with the given BCs, and, therefore, form
the complete orthonormalized set of basis functions. Any
Bose-symmetric WF of the variables $\textbf{r}_{1}, \ldots,
\textbf{r}_{N}$  for the Schr\"{o}dinger problem \textit{with
interatomic interaction} and periodic BCs can be expanded in this
basis. This is the ground for the theory of quantum liquids
constructed in \cite{yuv1,yuv2}. Hence, any WF $\Psi(\textbf{r}_{1},
\ldots, \textbf{r}_{N})$, being an eigenfunction of the momentum
operator of the system of $N$ identical bosons and corresponding to
the momentum $\hbar\textbf{p}$, can be presented in the form of a
sum
  \begin{eqnarray}
\Psi_{\textbf{p}}=a_{1}\psi^{f}_{\textbf{p}}+\sum\limits_{\textbf{k}_{1}\textbf{k}_{2}}^{\textbf{k}_{1}+\textbf{k}_{2}=\textbf{p}}
a_{2}(\textbf{k}_{1},\textbf{k}_{2})\psi^{f}_{\textbf{k}_{1}\textbf{k}_{2}}+
\ldots
+\sum\limits_{\textbf{k}_{1}\ldots\textbf{k}_{N}}^{\textbf{k}_{1}+\ldots
+\textbf{k}_{N}=\textbf{p}}
a_{N}(\textbf{k}_{1},\ldots,\textbf{k}_{N})\psi^{f}_{\textbf{k}_{1}\ldots\textbf{k}_{N}}.
 \label{psi-7}\end{eqnarray}
Using Eqs. (\ref{psi-1})--(\ref{psi-6}), this expansion can be written as
   \begin{eqnarray}
 \Psi_{\textbf{p}} & =&
  b_{1}(\textbf{p})\rho_{-\textbf{p}} +
 \sum\limits_{\textbf{q}_{1}\neq 0}^{\textbf{q}_{1}+\textbf{p}\neq 0}
  \frac{b_{2}(\textbf{q}_{1};\textbf{p})}{2!N^{1/2}}
 \rho_{\textbf{q}_{1}}\rho_{-\textbf{q}_{1}-\textbf{p}}
 + \ldots \nonumber \\ &+& \sum\limits_{\textbf{q}_{1},\ldots,\textbf{q}_{N-1}\neq 0}^{\textbf{q}_{1}+\ldots +\textbf{q}_{N-1}+\textbf{p}\not= 0}
  \frac{b_{N}(\textbf{q}_{1},\ldots,\textbf{q}_{N-1};\textbf{p})}{N!N^{(N-1)/2}}
 \rho_{\textbf{q}_1}\cdots\rho_{\textbf{q}_{N-1}}
 \rho_{-\textbf{q}_{1} - \ldots - \textbf{q}_{N-1}-\textbf{p}}.
       \label{psi-9}\end{eqnarray}
Here, the wave vectors $\textbf{k}_l$, $\textbf{q}_l$,
$\textbf{p}_l$, $\textbf{p}$ are quantized by the law (for 3D) $
\textbf{q}=2\pi \left (j_{x}/L_{x}, j_{y}/L_{y}, j_{z}/L_{z} \right
)$, where $j_{x}, j_{y}, j_{z}$ are integers, and  $L_{x}, L_{y},
L_{z}$ are the system sizes.

If GS is non-degenerate, then the wave function of GS is always
positive and corresponds to zero momentum (as was shown in Sect. 2).
Therefore, it can be presented in the form $\Psi_{0} = C\cdot
e^{S_{0}}$, where $S_{0}$ is $\Psi_{\textbf{p}}$ (\ref{psi-9}) with
$\textbf{p}=0$ \cite{yuv1,holes2020}:
\begin{eqnarray}
  S_{0}&=&
   \sum\limits_{\textbf{q}_{1}\neq 0}\frac{a_{2}(\textbf{q}_{1})}{2!}\rho_{\textbf{q}_{1}}
   \rho_{-\textbf{q}_{1}}+ \sum\limits_{\textbf{q}_{1},\textbf{q}_{2}\neq 0}^{\textbf{q}_{1}+\textbf{q}_{2}\not= 0}
  \frac{a_{3}(\textbf{q}_{1},\textbf{q}_{2})}{3!N^{1/2}}
 \rho_{\textbf{q}_{1}}\rho_{\textbf{q}_{2}}\rho_{-\textbf{q}_{1} - \textbf{q}_{2}}+\ldots +
   \nonumber\\  &+&
  \sum\limits_{\textbf{q}_{1},\ldots,\textbf{q}_{N-1}\neq 0}^{\textbf{q}_{1}+\ldots +\textbf{q}_{N-1}\not= 0}
  \frac{a_{N}(\textbf{q}_{1},\ldots,\textbf{q}_{N-1})}{N!N^{(N-2)/2}}
 \rho_{\textbf{q}_1}\ldots\rho_{\textbf{q}_{N-1}}
 \rho_{-\textbf{q}_{1} - \ldots - \textbf{q}_{N-1}}.
    \label{psi-10}   \end{eqnarray}
In this case, the constant $b_{1}(\textbf{0})\rho_{\textbf{0}}\equiv
\sqrt{N}b_{1}(\textbf{0})$ is taken into account in $C$. If each of
the functions $S_{j}$ from $S_{0}$ (\ref{1-2full}) is expanded in a
Fourier series in $j-1$ variables, then the resulting series can be
written in the form $S_{0}+const$ with $S_{0}$ (\ref{psi-10}). For
example, the following equality holds:
\begin{eqnarray}
&&\frac{1}{3!}\sum\limits^{j_{1}\neq j_{2},j_{3}; j_{2}\neq
j_{3}}_{j_{1}j_{2}j_{3}}S_{3}(\textbf{r}_{j_{1}}-\textbf{r}_{j_{2}},\textbf{r}_{j_{2}}-\textbf{r}_{j_{3}})=
\sum\limits_{\textbf{q}_{1},\textbf{q}_{2}\neq
0}^{\textbf{q}_{1}+\textbf{q}_{2}\not= 0}
  \frac{\tilde{a}_{3}(\textbf{q}_{1},\textbf{q}_{2})}{3!N^{1/2}}
 \rho_{\textbf{q}_{1}}\rho_{\textbf{q}_{2}}\rho_{-\textbf{q}_{1} - \textbf{q}_{2}}+
   \nonumber\\  &+&\sum\limits_{\textbf{q}_{1}\neq 0}\frac{g_{2}(\textbf{q}_{1})}{2!}\rho_{\textbf{q}_{1}}
   \rho_{-\textbf{q}_{1}}+\frac{1}{2!}\sum\limits^{j_{1}\neq
j_{2}}_{j_{1}j_{2}}\tilde{S}_{2}(\textbf{r}_{j_{1}}-\textbf{r}_{j_{2}})+const.
       \label{psi-13}\end{eqnarray}
The functions $\tilde{a}_{3}(\textbf{q}_{1},\textbf{q}_{2})$,
$g_{2}(\textbf{q}_{1})$, and
$\tilde{S}_{2}(\textbf{r}_{j_{1}}-\textbf{r}_{j_{2}})$ can be easily
determined, by expanding
$S_{3}(\textbf{r}_{j_{1}}-\textbf{r}_{j_{2}},\textbf{r}_{j_{2}}-\textbf{r}_{j_{3}})$
in a Fourier series in the variables
$\textbf{r}_{j_{1}}-\textbf{r}_{j_{2}}$,
$\textbf{r}_{j_{2}}-\textbf{r}_{j_{3}}$ and making simple
transformations.  This proves that if GS is non-degenerate and
$\ln{\Psi_{0}}$ can be expanded in a Fourier series, then
$\ln{\Psi_{0}}= S_{0}+const$, where $S_{0}$ has the form
(\ref{psi-10}) or the equivalent form (\ref{1-2full}).

Formula (\ref{psi-9}) can be written in the form (\ref{1-new0000}),
(\ref{psip-2}) (with other $b_{j}$), which is more suitable for the
description of quasiparticles.

We note that function (\ref{psi-10}) can be isotropic (formula
(\ref{1-2full})) or anisotropic (formula (\ref{1-5b})).  The above
analysis (in this appendix) does not allow us to establish whether
$S_{0}$ (\ref{psi-10}) is isotropic. However, it was shown in book
\cite{gilbert} (see Appendix 2 below) that GS of a Bose system is
non-degenerate. Moreover, we have shown in Section 2 that any
anisotropic state of the infinite 3D  Bose system is degenerate.
This property establishes a restriction for $S_{0}$ (\ref{psi-10}):
for the infinite 3D Bose system, this function must be isotropic,
since it corresponds to the ground state. Such function can depend
only on the quantities that are invariable at any  rotation, as well
as at any translations.

Therefore, we can write $S_{0}$ in the form
$S_{0}=S_{0}(f_{1},\ldots, f_{j})$, where $f_{l}$ is
$(\textbf{r}_{j_{1}}-\textbf{r}_{j_{2}})^{2}$,
$(\textbf{r}_{j_{1}}-\textbf{r}_{j_{2}})(\textbf{r}_{j_{2}}-\textbf{r}_{j_{4}})$,
or
$(\textbf{r}_{j_{1}}-\textbf{r}_{j_{2}})(\textbf{r}_{j_{3}}-\textbf{r}_{j_{4}})$.
It can be proved directly that such $S_{0}$ satisfies the equality
\begin{eqnarray}
\hat{\textbf{L}}S_{0}(\textbf{r}_{1},\ldots,\textbf{r}_{N})=0
       \label{psi-11}\end{eqnarray}
for all $N\geq 2$. Here, $\hat{\textbf{L}}$ is the operator of total
angular momentum of a system of $N$ particles:
\begin{eqnarray}
 \hat{\textbf{L}}=\sum\limits_{j=1}^{N} [\textbf{r}_{j}\times\hat{\textbf{p}}_{j}]
 =-i\hbar\sum\limits_{j=1}^{N} \left [\textbf{r}_{j}\times\frac{\partial}{\partial\textbf{r}_{j}}\right ].
       \label{psi-12}\end{eqnarray}
It is natural to expect that, at the transition to a \textit{finite}
periodic system, the structure of $S_{0}$ does not change. Hence,
for a finite periodic 3D system, $S_{0}$ should also be isotropic.
Thus, function (\ref{1-5}), (\ref{1-2full}) specifies the general
form of GS WF  of a periodic 3D Bose system.

\section{Appendix 2. Proof of the nondegeneracy of the ground state}
In the classical monograph by R. Courant and D. Hilbert
\cite{gilbert}, the theorem of nodes was proved for one particle
located in a finite two-dimensional volume with zero BCs. The proof
can be easily generalized to the case of a large number of particles
and any dimensionality of space. The proof in \cite{gilbert} admits
the presence of the degeneracy ($E_{j}=E_{j-1}$) of finite
multiplicity (multiplicity is finite, if the volume of the system is
finite, see \cite{gilbert}, $\S 2$). If GS is doubly degenerate,
then one of the states is described by a nodeless WF $\psi_{1}$
(according to the theorem of nodes). WF $\psi_{2}$ of the second
state can have a single node, according to the same theorem. On the
other hand, $\psi_{2}$ should be orthogonal to $\psi_{1}$ and,
therefore, \textit{must} have at least one node. Thus, the theorem
of nodes \cite{gilbert} admits the possibility for GS to be
degenerate.

The nondegeneracy of GS was proved at the other place of book
\cite{gilbert}. The proof is based on the Jacobi method (see
\cite{gilbert}, $\S 7$). We will give it in a slightly more detailed
form.

Consider the Schr\"{o}dinger equation
\begin{eqnarray}
   -\triangle \psi + U(x,y)\psi - E \psi=0
       \label{s1}\end{eqnarray}
for one particle located in a 2D region $G=(x,y)$ with the zero BCs
($\psi(x,y)=0$ on the boundary of the region $G$). Here, $U(x,y)$ is
a potential, and we set $\hbar=2m=1$. If there exists a solution
$\psi_{1}$ of Eq. (\ref{s1}) corresponding to the smallest
eigenvalue $E_{1}$, then $\psi_{1}$ can be found by solving the
following variation problem \cite{gilbert}: the inequality
\begin{eqnarray}
 D[\varphi]=\int\limits_{G}dx dy (\varphi_{x}^{2}+\varphi_{y}^{2}+U\varphi^{2})\geq  E_{1}\int\limits_{G}dx dy \varphi^{2}
       \label{s2}\end{eqnarray}
should be satisfied for all functions $\varphi(x,y)$ that are equal
to zero on the boundary of the region $G$ and have ``good''
properties ($\varphi$ should be continuous, whereas $\varphi_{x}$
and $\varphi_{y}$ should be piecewise continuous). Here,
$\varphi_{x}\equiv
\partial\varphi/\partial x$, $\varphi_{y}\equiv
\partial\varphi/\partial y$.  Inequality (\ref{s2})
becomes  an equality only for $\varphi(x,y)=c_{1}\psi_{1}(x,y)$,
where $c_{1}=const$. It follows from the theorem of nodes that
$\psi_{1}$ has no nodes \cite{gilbert}.

Assume that GS is degenerate and corresponds to two functions:
$\psi_{1}$ and $\psi_{2}$. In this case, $\psi_{1}$ has no nodes,
and $\psi_{2}$ must have one node (as was noted above). In this case
from the variation viewpoint, $\psi_{1}$ and $\psi_{2}$ satisfy
condition (\ref{s2}) and the zero BCs, and $\psi_{2}$ satisfies
additionally the condition of orthogonality of the functions
$\psi_{1}$ and $\psi_{2}$. Since $\psi_{1}$ has a constant sign
everywhere inside $G$, we may set
$\psi_{2}(x,y)=\vartheta(x,y)\psi_{1}(x,y)$. We will see now whether
such solution is possible. We set
$\varphi(x,y)=\eta(x,y)\psi_{1}(x,y)$ in $D[\varphi]$ (\ref{s2}).
Then
\begin{eqnarray}
 D[\varphi]=\int\limits_{G}dx dy [\psi_{1}^{2}(\eta_{x}^{2}+\eta_{y}^{2})+\eta^{2}(\psi_{1x}^{2}+\psi_{1y}^{2})+
 2\psi_{1}\psi_{1x}\eta\eta_{x}+
 2\psi_{1}\psi_{1y}\eta\eta_{y}+U\eta^{2}\psi_{1}^{2}].
       \label{s3}\end{eqnarray}
Let us use the relations  $2\eta\eta_{x}= (\eta^{2})_{x}$,
$2\eta\eta_{y}= (\eta^{2})_{y}$ and integrate the terms with
$\eta\eta_{x}$ and $\eta\eta_{y}$ by parts. We obtain two integrals
over the boundary which  are equal to zero due to the zero BCs. The
remaining terms give
\begin{eqnarray}
 D[\varphi]=\int\limits_{G}dx dy [\psi_{1}^{2}(\eta_{x}^{2}+\eta_{y}^{2})-\eta^{2}\psi_{1}\triangle \psi_{1}+
 U\eta^{2}\psi_{1}^{2}].
       \label{s4}\end{eqnarray}
Since $\psi_{1}$ satisfies Eq. (\ref{s1}) with $E=E_{1}$, formula
(\ref{s4}) is reduced to
\begin{eqnarray}
 D[\varphi]=\int\limits_{G}dx dy [\psi_{1}^{2}(\eta_{x}^{2}+\eta_{y}^{2})+
 E_{1}\eta^{2}\psi_{1}^{2}]\geq E_{1}\int\limits_{G}dx dy \varphi^{2}.
       \label{s5}\end{eqnarray}
The equality is obtained only for $\eta(x,y)=C=const$. Hence, the
wave function corresponding to the energy $E_{1}$ can have only the
form $const\cdot\psi_{1}(x,y)$. Therefore, the solution
$\psi_{2}(x,y)=\vartheta(x,y)\psi_{1}(x,y)$ with $\vartheta(x,y)\neq
const$ is impossible. This proves that the lowest level is always
non-degenerate. In this case, any excited state $\psi_{j>1}(x,y)$
can be degenerate (because $\psi_{j>1}(x,y)$ has nodes, and,
therefore, the representation $\varphi(x,y)=\eta(x,y)\psi_{j}(x,y)$
is inapplicable).

If we pass in all formulae from $x,y$ to
$\textbf{r}_{1},\ldots,\textbf{r}_{N}$, the reasoning conserves its
validity. Therefore, the conclusion about the nondegeneracy of GS is
true for systems with any $N$ and for any dimensionality of space.
The above analysis was performed for a finite system under the zero
BCs. We may expect that the main conclusion holds for any BCs  and
for infinite systems. Apparently, the above consideration can be
applied to infinite systems, if the zero BCs are set at infinity.

\section{Appendix 3. The third law of thermodynamics, properties of $^4$He,  Monte Carlo simulations}
We now consider the third principle of thermodynamics. Some
researchers believe that namely a crystal (rather than a liquid)
corresponds to the genuine GS of a system, since the crystal is more
ordered in the $\textbf{r}$-space and, therefore, should be
characterized by a lower entropy. However, we have seen in Sect. 2
that
in point of fact a liquid is characterized by a higher symmetry as
compared with a crystal.  Moreover, according to quantum statistics,
the entropy is determined by properties of a system in the space of
quantum states (not in the $\textbf{r}$-space). It is given by the
formula $S=k_{B}\ln(N(E))$ \cite{huang}, where $N(E)$ is the number
of states with energy close to $E$. To what is $N(E)$ equal for the
GS of a Bose crystal? Inequality (\ref{e00}) implies that many
liquid states with energy close to $E_{0}^{c}$ must exist. If we
take them into account, we get $N(E_{0}^{c})\gg 1$ and $S\neq 0$.
Therefore, it is necessary to introduce the natural postulate: in
the calculations of thermodynamic quantities, one needs to take only
states of the phase under consideration (gas, liquid, or crystal)
into account in the statistical sum. In addition, the complete set
of eigenfunctions of the Hamiltonian should contain the solutions
for crystal lattices of various types and various spatial
orientations (see Subsect. 2.2). Of course, while describing a
crystal, it is necessary to consider in the statistical sum only the
states associated with one type  and one orientation of the lattice.
Then for each phase we obtain $N(E)= 1$ and $S= 0$ at $T=0$, i.e.,
the Nernst theorem is satisfied. However, if the lowest state of a
real finite crystal is $j$-fold degenerate, we obtain for it
$S(T=0)= k_{B}\ln(j)$.

$^4$He has particular properties. According to experiments, liquid
$^4$He (He II) at $P\approx 25\,atm$ and $T\lsim 0.8\,K$ solidifies
and transforms into a hcp crystal \cite{guyer}. In this case, for
liquid and solid $^4$He, we have, respectively,
$E_{0}^{l}\approx-6.6\,K$ \cite{abraham1970,roach1970} and
$E_{0}^{c}\approx-5.96\,K$ \cite{edwards1965}. That is, the GS
energy of a crystal by $0.6\,K$ \textit{higher}, than  $E_{0}$ of a
liquid. In this case, liquid and solid  helium have densities of
$0.1725\,g/cm^{3}$ \cite{abraham1970,grilly1973} and
$0.191\,g/cm^{3}$ \cite{edwards1965,grilly1973}, respectively. To
verify the basic inequality (\ref{e00}), we need to compare $E_{0}$
of liquid and solid helium at the same $\rho$. We can determine
$E_{0}$ of liquid helium at $\rho=0.191\,g/cm^{3}$ by the known
formula in \cite{abraham1970,roach1970}. In this case, one needs to
know $P(\rho)$ of He II at $\rho=0.1725$--$0.191\,g/cm^{3}$.
However, such data are not available, since He II does not exist at
such densities: it solidifies. It is significant that GS of liquid
helium at the crystallization pressure ($P\approx 25\,atm$) has a
lower energy, than GS of a crystal. Nevertheless, liquid helium
crystallizes. The reason is known and is as follows. $^4$He
possesses large zero oscillations. Therefore, at low pressures, the
crystal is unstable. As a result, the system at low pressures and
temperatures is in the state of underliquid. This is He II. As the
pressure increases, the ratio of the amplitude of zero oscillations
to the lattice period decreases \cite{guyer,pomer}. At $P\gsim
25\,atm$, the crystal embryos become stable and liquid $^4$He
crystallizes \cite{keesom,mendelssohn}. In this case, the formation
of microcrystals and the external pressure make the liquid state
unstable: the external pressure compresses the system, performs the
work, and increases the energy of the system up to $E_{0}$ of a
crystal. This results in the formation of a crystal. By such a
scenario, the ground state of $^4$He at $P> 25\,atm$ corresponds to
a liquid, but this state cannot be obtained. However, the
crystallization pressure should rise above $25\,atm$ provided we
prevent the formation of crystal nuclei (one needs to purify helium
from impurities and use the smooth walls with a microstructure
different from the structure of a helium crystal). We do not know
whether attempts to obtain high-density liquid helium in this way
were made before. If  He II with $\rho=0.1725$--$0.191\,g/cm^{3}$
could be obtained, it would be possible to verify inequality
(\ref{e00}) for $\rho=0.191\,g/cm^{3}$.

The analytic analysis of crystalline solutions is very complicated.
Therefore, the majority of theoretical studies of quantum crystals
were executed numerically by the Monte Carlo (MC)  method (see works
\cite{levesque1968,cep1993,cep1976,cep1978,mcmillan1965,reatto2009,vitiello1988,reatto1998,reatto2011,loubeyre1988,kalos1974,whitlock1979,cep1984,w87,sarsa2000,galli2003,moroni2000,loubeyre1986,b7,b11,l21}
and reviews \cite{cazorla2017,reatto1995,whitlock2006,ceperley1992};
an introduction to the MC methods can be found in books
\cite{koonin1986,gould2007}). Some MC methods are ``exact'' and are
independent of (by authors' opinion) a trial function
\cite{cep1993,cep1976,reatto2009,reatto2011,kalos1974,whitlock1979,w87,sarsa2000,galli2003,moroni2000,b7,b11,l21}.
However, the ``exact'' MC simulations do not give an analytic
solution for WF. Therefore, the authors believed that the solution
corresponds to a nodeless WF of a crystal,  by basing on indirect
signs. Usually, such sign is simply the density of the system equal
to the experimental density of the crystal
\cite{cep1993,mcmillan1965,vitiello1988,reatto1998,cep1984,sarsa2000,moroni2000}.
Sometimes, the authors distinguished a liquid solution and a
crystalline one by trial WFs
\cite{cep1993,cep1976,cep1978,whitlock1979,w87}, the inequality
$E_{0}^{c}(\rho)<E_{0}^{l}(\rho)$ \cite{levesque1968,kalos1974}, a
sharp turn on the curve $\langle r^{-12}\rangle(a_{1})$ ($a_{1}$ is
the parameter of the Bijl--Jastrow WF) \cite{mcmillan1965},  and a
jump of $\rho$ \cite{loubeyre1986,wood1957}. The basic property
allowing one to separate a quantum crystal from a liquid, namely,
the anisotropy of solutions, was studied only in a few works
\cite{reatto2009,reatto2011,whitlock1979,galli2003,b7,b11,l21}
(solutions for the classical system of hard balls see in
\cite{krauth2015}).

By means of the symmetry analysis we have shown in Sect. 2 that the
genuine GS must correspond to a liquid for any density. The symmetry
analysis is a more strong argument, than a MC simulation. Since the
former presents the exact information, whereas the latter gives only
an approximate one. Therefore, we suppose that the MC simulations
gave always a nodeless liquid solution or a crystalline solution
corresponding to a WF with nodes (instead of a nodeless crystalline
solution). In particular, the \textit{isotropic} function
$g_{2}(\textbf{r}_{1}-\textbf{r}_{2})$  was obtained for a
crystalline solution \cite{whitlock1979}. The authors of work
\cite{whitlock1979} interpreted such solution as a crystalline one,
only because it corresponds to the densities at  which the natural
helium is a solid substance. As for the isotropy, they related it to
large zero oscillations \cite{whitlock1979}. In our opinion, the
isotropy of $g_{2}(\textbf{r}_{1}-\textbf{r}_{2})$ indicates clearly
the liquid character of the solution. With the help of the shadow
PIGS (path integral ground state) MC method, the authors of work
\cite{galli2003} obtained a crystal-like distribution of atoms.
Within an analogous method the solutions  for 2D and 3D crystals
were obtained in \cite{reatto2011} and \cite{reatto2009},
respectively. The crystal character of a solution was determined by
the Bragg peak of the structural factor $S(k_{y})$ \cite{reatto2011}
or $S(k)$ \cite{reatto2009}. We think that works
\cite{reatto2009,reatto2011,galli2003,b7,b11,l21} present the
crystalline solutions corresponding to a local crystalline energy
minimum (see Fig. 1) and WFs with nodes.  As far as we understand,
the Green's function MC \cite{whitlock1979,w87}, PIGS MC
\cite{reatto2009,reatto2011,galli2003}, PIMC \cite{b7,b11}, and a
diffusion MC \cite{l21} are ``exact'' methods, but they do not allow
one to reliably clarify whether the obtained WF has nodes. In works
\cite{reatto2009,reatto2011}, no comparison of the parameters of the
Bragg peak and the lattice parameters was made. It is worth to
verify whether the lattice is one-dimensional (this would explain
the  smallness of the quantity $|E_{0}^{l}-E_{0}^{c}|\sim (0.001
\div 0.01)|E_{0}^{l }|$ \cite{cep1993,reatto2011}; indeed, the 1D
lattice corresponds to WF with a small number of nodes; therefore,
the energy has to be close to the energy of a nodeless liquid
solution). On the other hand, a small value of
$|E_{0}^{l}-E_{0}^{c}|$ can be connected with the Yukawa potential
\cite{cep1993}.

Usually, the MC simulations lead to the crystalline solution at a
sufficiently high concentration $n$. This is apparently because an
increase in many-particle corrections in (\ref{1-2full}) with $n$
\cite{yuv1,mt2006}. In this case, the trial two-particle function
$\Psi_{0}$ deviates more and more from the exact one. Therefore, the
difference between the trial $E_{0}$ and the exact one can exceed
the exact difference $|E_{0}^{l}-E_{0}^{c}|$. As a result, the
method can lead to the crystalline minimum region (Fig. 1). If the
system falls in the crystalline minimum in the process of
simulation, one can try to take it off this minimum by increasing
the step of simulation by one order of magnitude or by  starting a
new simulation with a small step using the previous or new trial
function. The ``exact'' MC simulations usually give information that
is insufficient to reliably determine the lattice type and to
clarify whether the crystal WF is nodeless. For the determination of
a type and dimensionality of the lattice, one needs to find the
function $g_{2}(\textbf{r}_{1}-\textbf{r}_{2})$ or $S(\textbf{k})$.
In this case, it is necessary to check $\Psi_{0}$ for nodes. We do
not know whether this can be realized with a good accuracy at $N>10$
(though the fixed-node MC methods allow one to determine, in
principle, the node structure of WF at $N\gg 1$ \cite{raja2001}). We
hope that the modern capabilities of computers and theory are
sufficient to perform the detailed studies and to clarify the
general picture with the help of MC simulations.

\section{Appendix 4. Formation of nuclei}
The theory of formation of nuclei of a new phase is not completed
(especially, the theory of crystallization), but its general
contours are apparently clear (see reviews \cite{sosso2016,fux1935}
and books
\cite{gibbs,strickland,frenkel,volmer,kuznetsov,danilov,chalmers,flemings,skripov,kashchiev,kelton}).
The nuclei of the other phase can be created on the walls of a
vessel and in bulk. We will consider only the  simpler bulk case.

The bulk condensation of a gas into a crystal or liquid occurs under
the avalanche-like increase in the number of nuclei of the new
phase. Such growth is possible, if $P$ or $T$ differs from the value
$P=P_{\infty}$ or $T=T_{\infty}$ corresponding to the condition of
equilibrium (\ref{ee-clc}). Consider a gas at low $P=P_{\infty}$ and
low $T=T_{\infty}$. Let us compress it isothermally so that the
pressure increases up to some $P> P_{\infty}$. In such gas, the
embryos of the liquid phase (microdrops) and the crystalline one
(microcrystals) should randomly appear. In a supersaturated gas
(vapor) at the pressure $P_{r}\geq P_{\infty},$ a droplet is in
equilibrium with a gas, if its radius $r$ satisfies the Kelvin's
formula \cite{huang,frenkel,volmer}:
\begin{eqnarray}
P_{r}(T)= P_{\infty}(T) \exp{\left (\frac{2\alpha_{lg} }{n
k_{B}T}\frac{1}{r}\right )},
       \label{pr}\end{eqnarray}
where  $n$ is the concentration of atoms in a droplet, $P_{\infty}$
is the saturated vapor pressure, $\alpha_{lg}$ is the coefficient of
surface tension of a liquid on the boundary with a gas. Let the
pressure $P_{r}$ correspond to the radius $r=r_{0}$, according to
(\ref{pr}). Then the droplets of radius $r<r_{0}$ must evaporate.
The condensation of atoms of a gas on a droplet decreases the
pressure in a gas, which makes it possible for the droplets of
radius $r> r_{0}$ to exist. As a result, the mean radius of droplets
must increase with the time, until the whole gas transforms into one
large drop \cite{huang}. The crystal embryos in a gas can be
described analogously. We will get formula (\ref{pr}), where the
parameters of a microdrop should be replaced by those of a
microcrystal.

According to a more detailed theory, the process of formation of
embryos is as follows
\cite{frenkel,fux1935,volmer,chalmers,flemings,skripov,kashchiev}.
The fluctuations in a gas result in the spontaneous formation of
microscopic embryos of a liquid (microdrops) and a crystal
(microcrystals) in a gas. The embryo can randomly capture atoms of
the gas, which will lead to the growth of this embryo. The reverse
process is possible as well. As a result, some (non-stationary,
generally speaking) distribution of embryos over sizes should be
formed. In this case, the embryos of sizes larger than the critical
one (Eq. (\ref{pr})) must unboundedly grow. Such embryos are usually
called nuclei. If the system is supplied with a gas in the amount
compensating the loss due to the formation of nuclei, we get a
stationary distribution of nuclei and the continuous transformation
of small nuclei into large ones. If such pumping of a gas is absent,
then in the usual case (isothermal formation of nuclei in a closed
system with permanent supersaturation) the non-stationary process
eventually becomes stationary \cite{skripov,kashchiev,kelton}.
Therefore, we may consider the process to be stationary. The kinetic
analysis shows that, in this case, the rate $J$ of homogeneous
(i.e., without exterior impurities) formation of nuclei is
\cite{sosso2016,strickland,frenkel,volmer,danilov,chalmers,flemings,skripov,kashchiev,kelton}
\begin{eqnarray}
J= n_{g}Be^{-\frac{W}{k_{b}T}},
       \label{j2}\end{eqnarray}
where $n_{g}$ is the gas concentration, $B$ is the kinetic factor
(which can depend on $P$ and $T$), $W> 0$ is the work of formation
of a critical nucleus (an embryo of such size for which $W$ is
maximum at the given $P$ and $T$). Condition (\ref{pr}) yields the
radius of such a nucleus as a function of $P=P_{r}$ at
$T=T_{\infty}=const$. Frequently, the dependence of the critical
radius $r$ on $T$ at $P=P_{\infty}=const$ is studied. Then
\cite{frenkel,chalmers}
\begin{eqnarray}
r=\frac{2\alpha_{lg}T_{\infty}}{n q(T_{\infty}-T)},
       \label{tr}\end{eqnarray}
where $n$ is the concentration of atoms in a nucleus,
$q=T[s_{g}(P,T)-s_{l,c}(P,T)]$ is the latent heat of the phase
transition per atom. As is seen, the higher the supercooling of a
vapor, the less the nucleus radius.

It is difficult to calculate the value of $B$ in (\ref{j2}).
Different models give different values. Within the classical
approach (high $T$ and large nuclei), J. Gibbs \cite{gibbs}  found
$W$ for a critical  liquid nucleus,
\begin{eqnarray}
W_{l}= \varsigma \alpha_{lg} /3,
       \label{ww}\end{eqnarray}
and for a critical crystal  nucleus,
\begin{eqnarray}
W_{c}= \sum\limits_{j}\varsigma_{j} \alpha_{j} /3.
       \label{ww2}\end{eqnarray}
Here, $\varsigma=4\pi r^{2}$ is the droplet surface area, $j$ is the
number of a crystal face, $\varsigma_{j}$ is the area of the $j$-th
face of a crystal, and $\alpha_{j}$ is the coefficient of surface
tension for the $j$-th face of the crystal which contacts with the
gas. It is useful to write formula (\ref{ww2}) in the form
\cite{strickland}
\begin{eqnarray}
W_{c}= \varsigma \bar{\alpha}_{cg}/3,
       \label{ww3}\end{eqnarray}
where $\bar{\alpha}_{cg}$ is the average coefficient of surface
tension of the crystal on the boundary with the gas, and $\varsigma$
is the area of a sphere with the volume equal to that of a
crystalline nucleus. At $T=T_{\infty}$ and $P=P_{\infty},$ the
radius of a critical nucleus is $r=\infty$. Therefore, $J$ turns to
zero, which corresponds to the equilibrium of phases.

Apparently, the underliquid can be obtained easier by means of the
isothermal compression of a gas, than by its isobaric cooling. We
now consider only the first way. The modern theory cannot exactly
conclude whether  the compressed gas will turn into a liquid or a
crystal. This is not surprising, because the process of transition
of one phase into another one is complex and depends on many
factors.

First, we note that $W$  is less at the condensation of a gas on the
surface, than at the condensation in bulk
\cite{fux1935,volmer,kuznetsov,flemings,kashchiev}. In particular,
the work of formation of a dome-shaped critical nucleus (liquid or
crystalline) of radius $r_{0}$ on a solid wall is
\cite{volmer,kuznetsov,flemings,kelton}
\begin{eqnarray}
W^{2D}= (\varsigma \alpha /12)[2+\cos{\theta}][1-\cos{\theta}]^{2},
       \label{ww2d}\end{eqnarray}
where $\varsigma=4\pi r_{0}^{2}$,  $\theta$ is the angle between the
nucleus surface and the wall, $\alpha$ is the surface tension of the
nucleus that is in contact with a gas. At the complete nonwetting
($\theta=\pi$), the value of $W^{2D}= \varsigma \alpha /3$ coincides
with that of the bulk work $W$ (\ref{ww}) or (\ref{ww3}). If the
wetting is present ($\theta<\pi$), then $W^{2D}<W_{l}, W_{c}$, and a
nucleus can be easier formed on the wall, than in bulk. Of course,
crystallization is a complex process that is not reduced to the
formation of dome-like nuclei. However, formula (\ref{ww2d}) shows
that a crystal nucleus can be easier formed on the surface, than in
bulk \cite{strickland,volmer,chalmers,flemings,kashchiev}.
Therefore, if the gas contains solid impurities (or the walls of a
vessel contain some inhomogeneities able to become the centers of
condensation), then the surface condensation, rather than the bulk
one, is realized. In practice, the impurities and inhomogeneities of
walls are usually present. Moreover,  $W^{2D}_{l}<W_{l}$ even for
the ideally smooth wall provided $\theta<\pi$. Therefore, the
condensation of a gas usually occurs on the walls or on impurity
particles.

According to experiments, at $T<T_{3}$ a gas condenses into a
crystal. This is because the gas-crystal curve lies below the
gas-liquid one (see Fig. 2). Microscopically, this means that
$W^{2D}_{c}$ corresponding to the formation of a two-dimensional
critical crystal nucleus is less than the work $W^{2D}_{l}$ of the
formation of an analogous liquid nucleus. The reason for this is
that the crystalline structure of a substrate usually decreases
$W^{2D}_{c}$ and thus stimulates the formation of namely crystalline
nuclei. In particular, the condensation of a gas into a crystal
becomes more intense, if a substrate on which the condensation
occurs is a crystal of a close structure
\cite{fux1935,volmer,danilov,flemings,kashchiev}, because in this
case $W^{2D}_{c}$ decreases.

In practice, the formation of crystal nuclei can be prevented  if
the gas is well purified from impurity particles and the vessel with
very smooth walls is used. In addition, the microstructure of walls
of a vessel should be significantly different from the
microstructure of a crystal, into which the gas can condense. Under
these conditions, the condensation of a gas into a liquid (on the
walls or in bulk) should be dominant.

Assume that the bulk homogeneous mechanism of spontaneous formation
of nuclei is realized. In this case, crystalline and liquid nuclei
will arise. The rate of each of these processes is given by formula
(\ref{j2}), where $W$ is determined by formulae (\ref{ww}) or
(\ref{ww3}). It is clear that $J_{c}\ll J_{l}$ at $T\rightarrow 0$
provided
\begin{eqnarray}
\lambda \equiv W_{c}/W_{l} > 1.
       \label{lam}\end{eqnarray}
In this case, the condensation of a gas into droplets is more
probable. Let us find the conditions under which relation
(\ref{lam}) is satisfied. Formulae (\ref{pr}), (\ref{ww}), and
(\ref{ww3}) yield
\begin{eqnarray}
\lambda
=\frac{\bar{\alpha}^{3}_{cg}}{\alpha^{3}_{lg}}\frac{n^{2}_{l}}{n^{2}_{c}}
\frac{[\ln{(P/P^{lg}_{\infty}}]^{2}}{[\ln{(P/P^{cg}_{\infty}}]^{2}},
       \label{lam2}\end{eqnarray}
where $n_{l}$ and $n_{c}$ are the concentrations of atoms in a
microdrop and a microcrystal, respectively, at the same pressure
$P$. We set $P^{lg}_{\infty}=\zeta \cdot P^{cg}_{\infty}$,
$\bar{\alpha}_{cg}=(1+\eta)\alpha_{lg}$, $n_{c}=(1+\vartheta)n_{l}$,
and  $(1+\eta)^{3}(1+\vartheta)^{-2}=(1+\phi)^{2}$. Here,
$P^{lg}_{\infty}$ and $P^{cg}_{\infty}$ are the equilibrium
pressures on the gas-liquid and gas-crystal curves, respectively. As
a rule, $|\eta |, |\vartheta | \ll 1$. Therefore, $|\phi|\ll 1$ as
well. Relation (\ref{lam2}) implies that inequality (\ref{lam})
holds at
\begin{eqnarray}
P/P^{cg}_{\infty} > \zeta^{\frac{1+\phi}{\phi}}.
       \label{lamp}\end{eqnarray}
That is, at $T\rightarrow 0$  the rate of formation of liquid nuclei
is much higher than that for crystal nuclei, if the gas is
isothermally compressed at a pressure $P$ exceeding
$P^{cg}_{\infty}$ by $\zeta^{\frac{1+\phi}{\phi}}$ times. The
quantity $\bar{\alpha}_{cg}$ can be estimated in the following way.
By the rate of formation of crystal nuclei  in a liquid, we can find
$\bar{\alpha}_{cl}$: usually, $\bar{\alpha}_{cl}\approx (0.1 \div
0.2) \alpha_{gl}$ (for temperatures close to the melting one; see
Table III.1 in \cite{strickland}). It is natural to assume that
$\bar{\alpha}_{cg}=\alpha_{gl}+\tilde{c}\bar{\alpha}_{cl}$, where
$\tilde{c}\simeq -1$, if the density of a crystal is less than that
of a liquid, and $\tilde{c}\simeq 1$ in the opposite case. For most
substances, the crystal is denser than the liquid ($\vartheta \simeq
0.1$). Therefore, we expect for them that
$\bar{\alpha}_{cg}\approx\alpha_{gl}+\bar{\alpha}_{cl}$, i.e., $\eta
\simeq 0.15$.  However, for some substances (e.g., ice) $\eta $ and
$\vartheta$ are significantly different and can be negative. For the
characteristic values $\eta = 0.15$ and $\vartheta =0.1,$ we get
$\phi \approx 0.1$, and (\ref{lamp}) gives $P/P^{cg}_{\infty} >
\zeta^{11}$. For the inert elements, the triple point corresponds to
$P_{3}\sim 1\,atm$. Therefore, at $T\ll T_{3}$ we have
$P^{cg}_{\infty} \ll 1\,atm$. According to the analysis in Sect. 3,
at $T\ll T_{3}$ the value of $\zeta$ is close to 1. Therefore, the
pressure $P
> \zeta^{11} P^{cg}_{\infty}$ at which the gas should condense
into droplets is quite achievable.

It was asserted in some works \cite{volmer,kashchiev} that, for the
vapor--crystal and vapor--liquid transitions, one needs to set
$B=B^{\prime}e^{Cq/(k_{b}T)}$ in formula (\ref{j2}). Here, the
constant $C$ depends on the mechanism ($|C|\simeq 1$), $q$ is the
latent heat of sublimation or evaporation, and $B^{\prime}$ may
slightly depend on $T.$ Above, we neglected the factor
$e^{Cq/(k_{b}T)}$. This is justified, if the phase transition occurs
at a not too high supersaturation (in this case, the critical radius
$r$ is large, and, therefore, $W\gg |C|q$).

For the surface mechanism of formation of nuclei, the formulae are
significantly more complicated, especially for crystalline nuclei.
In the last case, the work $W_{c}$ depends also on the relationship
of the crystalline structures of a nucleus and the substrate
\cite{sosso2016,fux1935,volmer,danilov,flemings,kashchiev}. We did
not make estimates for this case. Most likely, the ratio
$P/P^{cg}_{\infty}$ is not too different from (\ref{lamp}).
Therefore, if the microstructures of the wall and crystal nuclei are
strongly different and the wall is very smooth, we may expect that
at the pressure $P> (2\div 3)P^{cg}_{\infty}$ the surface formation
of liquid nuclei is more probable, than the surface formation of
crystal nuclei. In this case, the gas should condense into a liquid
when compressed. Moreover, if atoms of a gas interact weakly with
atoms of the walls, then the bulk formation of nuclei (drops or
crystals) should be more intense, as compared with the formation of
nuclei on the walls.

Our analysis is rather crude, but the main conclusions are
apparently qualitatively right. Thus, the experiment on gas
compression should be carried out with different walls of vessels at
several different temperatures $T\ll T_{3}$. The condensation of a
gas into a liquid has to be more probable than the crystallization,
provided that (i) the microstructures of the wall and crystal nuclei
are significantly different (or atoms of the gas interact weakly
with atoms of the wall and relation (\ref{lamp}) holds) and (ii) the
gas is purified from impurities. Condition (i) can be fulfilled by
covering the  internal surfaces of the walls of a vessel with a
solid amorphous substance \cite{espinosa2019,sosso2016} or with a
microlayer of helium-II. Perhaps, this is the simplest way to obtain
the underliquid.


       \end{document}